\documentclass[prd,reprint,superscriptaddress,preprintnumbers,nofootinbib,amsmath,amssymb,aps]{revtex4-2}

\usepackage{ads}
\usepackage{graphicx}
\usepackage{dcolumn}
\usepackage{color}
\usepackage[colorlinks=true,citecolor=blue,urlcolor=blue]{hyperref}

\begin{document}

\title{What does cosmology tell us about the mass of thermal--relic dark matter?}
\author{Rui An}
\thanks{Email: anrui@usc.edu}
\author{Vera Gluscevic}
\thanks{Email: gluscevi@usc.edu}
\affiliation{Department of Physics and Astronomy, University of Southern California, Los Angeles, CA 90089, USA}
\author{Erminia Calabrese}
\affiliation{School of Physics and Astronomy, Cardiff University, The Parade, Cardiff, Wales CF24 3AA, UK}
\author{J. Colin Hill}
\affiliation{Department of Physics, Columbia University, New York, NY 10027, USA}
\affiliation{Center for Computational Astrophysics, Flatiron Institute, New York, NY 10010, USA}

%%%%%%%%%%%%%%%%%%%%%%%%
\begin{abstract}

The presence of light thermally coupled dark matter affects early expansion history and production of light elements during the Big Bang Nucleosynthesis. Specifically, dark matter that annihilates into Standard Model particles can modify the effective number of light species in the universe $N_\mathrm{eff}$, as well as the abundance of light elements created buring BBN. These quantities in turn affect the cosmic microwave background (CMB) anisotropy. We present the first joint analysis of small-scale temperature and polarization CMB anisotropy from Atacama Cosmology Telescope (ACT) and South Pole Telescope (SPT), together with \textit{Planck} data and the recent primordial abundance measurements of helium and deuterium to place comprehensive bounds on the mass of light thermal--relic dark matter. We consider a range of models, including dark matter that couples to photons and Standard-Model neutrinos. We discuss the sensitivity of the inferred mass bounds on measurements of $N_\mathrm{eff}$, primordial element abundances and the baryon density, and quantify the sensitivity of our results to a possible existence of additional relativistic species. We find that the combination of ACT, SPT, and \textit{Planck} generally leads to the most stringent mass constraint for dark matter that couples to neutrinos, improving the lower limit by 40\%--80\%, with respect to previous \textit{Planck} analyses. On the other hand, the addition of ACT and SPT leads to a slightly weaker bound on electromagnetically coupled particles, due to a shift in the preferred values of $Y_\mathrm{p}$ and $N_\mathrm{eff}$ driven by the ground based experiments. In most scenarios, the combination of CMB data has a higher constraining power than the primordial abundance measurements alone, with the best results achieved when all data are combined. Combining all CMB measurements with  primordial abundance measurements, we rule out masses below $\sim$4 MeV at 95\% confidence, for all models.
We show that allowing for new relativistic species can weaken the mass bounds for dark matter that couples to photons by up to an order of magnitude or more.
Finally, we discuss the reach of the next generation of the CMB experiments in terms of probing the mass of the thermal relic dark matter.
\end{abstract}
%%%%%%%%%%%%%%%%%%%%%%%%

\maketitle

%%%%%%%%%%%%%%%%%%%%%%%%
\section{Introduction}

A variety of astrophysical and cosmological observations indicate that a significant fraction of the matter in our Universe is composed of dark matter (DM) \cite{2009GReGr..41..207Z, 1970ApJ...159..379R, 2003ARA&A..41..645R, 2006ApJ...648L.109C, 2020A&A...641A...6P}. Despite many decades of dedicated searching, the nature of DM remains a mystery and exploring its essence is one of the most challenging tasks for fundamental physics today.
Some of the most compelling candidate models invoke DM that is in thermal equilibrium with the Standard--Model (SM) particles in the early universe \cite{1965JETP...21..656Z,1965PhL....17..164Z,1966PhRvL..17..712C,2005PhR...405..279B}. Among the thermal--relic models are the Weakly Interacting Massive Particles (WIMPs), the main focus of many current direct detection experiments \cite{2018EPJC...78..203A, 2018RPPh...81f6201R}. 
While WIMPs were originally considered to have masses in the GeV--TeV range, current null results from direct detection searches have inspired compelling WIMP-like and other models for light thermal--relic DM, with sub-GeV masses \cite{2008PhRvD..77h7302H, 2008PhRvL.101w1301F, 2007PhRvD..76j3515H}. In this work, we consider the most general observable consequences of light thermal--relic DM, including the effects on the expansion history and the Big Bang Nucleosynthesis (BBN), and infer comprehensive bounds on its particle mass $m_\chi$, using all available data. 

BBN took place in the very early universe, beginning a fraction of a second after the Big Bang, and ending tens of minutes later. During this time, masses and interactions of particles present in the primordial plasma have defined the expansion history and thermal history, defining the abundances of light elements created during BBN.
In particular, if DM particles are in thermal contact with the rest of the plasma during BBN (as is the case for majority of the popular WIMP models in current literature \cite{1996PhR...267..195J, 2012JCAP...12..027H, 2012AnP...524..479B, 2013JCAP...08..041B, 2013PhRvD..87j3517S, 2014PhRvD..89h3508N, 2015PhRvD..91h3505N, 2016PhRvD..94j3525W, 2019JCAP...02..007E, 2019JCAP...04..029D, 2020JCAP...01..004S}), and if their masses are around $\sim$0.01--20 MeV, they become non-relativistic right around the time of BBN. The resulting DM annihilation into SM particles can affect the early expansion history and the production of light elements \cite{1986PhRvD..34.2197K, 2004PhRvD..70d3526S, 2004JPhG...30..279B}, including the abundance of helium--4, $Y_\mathrm{p}$, and deuterium, $Y_\mathrm{D}\equiv(D/H) \times 10^5$. Annihilation products can additionally alter the radiation content in the universe, changing the effective number of light species, $N_\mathrm{eff}$.
These effects are captured by the cosmic microwave background (CMB) anisotropy and can be directly traced by measurements of primordial abundances of light elements in Lyman-$\alpha$ forest systems.

We focus on inferring the lower bound on $m_\chi$ from both these data sets, for a minimal set of modeling assumptions, requiring only that DM is in thermal equilibrium with the rest of the universe prior to BBN. In this context, we consider two general scenarios: one in which DM couples electromagnetically to the SM, and another in which it only couples to the SM neutrinos. In both scenarios, the light DM annihilates and transfers its energy and entropy to the remaining SM particles directly \cite{1986PhRvD..34.2197K,2013PhRvD..87j3517S,2014PhRvD..89h3508N,2015PhRvD..91h3505N}. Within each scenario, we consider four types of DM spin statistic: real scalar, complex scalar, Majorana Fermion, and Dirac Fermion.  Notably, we do \textit{not} attempt to relate the cosmological effects of $m_\chi$ to the strength of the coupling to the SM; provided that the coupling is sufficient to ensure thermal equilibrium prior to BBN, the effects of $m_\chi$ are \textit{independent} on details of the interaction model. This choice means that we do \textit{not} provide a connection between the relic abundance of DM and its mass, nor do we consider late--time DM annihilation constraints---these are additionally dependent on the strength and type of the interaction at hand, and we leave such considerations for future work. 

We note that the CMB constraints on the mass of light DM have been previously derived using \textit{Planck} data and primordial element abundance measurements (e.~g.~see \cite{2013JCAP...08..041B, 2014MmSAI..85..175S, 2014PhRvD..89h3508N, 2015PhRvD..91h3505N, 2019JCAP...02..007E, 2021arXiv210903246G}). In this work, in addition to the latest \textit{Planck} measurements \cite{2020A&A...641A...6P}, we include the most recent public releases of the CMB small-scale measurements from the Atacama Cosmology Telescope (ACT) \cite{2020JCAP...12..047A, 2020JCAP...12..045C} and South Pole Telescope (SPT) \cite{2021PhRvD.104b2003D}, as well as the primordial element abundance measurements from \cite{2020PTEP.2020h3C01P}. ACT and SPT in particular provide higher resolution measurements of the polarization anisotropy, increasing the sensitivity to small angular scales, and thus enabling complementary constraints on cosmological parameters to \textit{Planck} data. Since a primary effect of $m_\chi$ on the CMB is through $N_\mathrm{eff}$, we further extend the analyses by allowing for additional light degrees of freedom to exist in the universe. We quantify the impact of this additional freedom on the inferred mass bounds. 

We find that the DM mass bounds can have a notable dependence on the choice of the data set and assumptions about other light degrees of freedom. In particular, the inclusion of ACT and SPT data leads to slightly different preferred values of $Y_\mathrm{p}$ and $N_\mathrm{eff}$, as compared to \textit{Planck} alone, resulting in an improved mass limit for neutrino coupled DM and a slightly less stringent bound on electromagnetically coupled DM mass. Combining all CMB measurements with the measurements of helium and deuterium abundance, we infer the lower mass limit of $\sim$4 MeV at $95\%$ confidence level (CL), regardless of the model. We further show that allowing for the existence of new relativistic species, parameterized by the effictive number of additional (neutrino--like) degrees of freedom $\Delta N_{\nu}$, the mass bounds are significantly less stringent for electromagnetically coupled models. We discuss the implications of possible systematic effects on the reported mass inference. Finally, we show that the future--generation ground--based CMB measurements from the Simons Observatory \cite{2019JCAP...02..056A} and CMB-S4 \cite{2016arXiv161002743A, 2019arXiv190704473A} will saturate the precision of CMB mass bounds.

The organization of the paper is as follows. In Section \ref{sec:BBN}, we briefly review the most relevant physics behind the standard BBN (SBBN) model, and discuss the alterations of SBBN that include light thermal--relic DM. 
In Section \ref{sec:CMB}, we quantify the impact of light DM on the CMB power spectrum. 
In Section \ref{sec:constraints}, we present the current observational data sets, our analysis method, and the resulting constraints on DM mass. In Section \ref{sec:forecasts}, we show the projected sensitivity of the upcoming CMB experiments. Finally, we summarize and discuss our findings in Sec.~\ref{sec:summary}. 

%%%%%%%%%%%%%%%%%%%%%%%%%%%%%%%%%%%%%%%
\section{Effects on primordial abundances}\label{sec:BBN}

Within the standard models of particle physics and cosmology, the universe was radiation-dominated during BBN. At the beginning of this process, it contained electrons and positrons (denoted here as $e^{\pm}$), photons ($\gamma$), three neutrino species ($\nu$), a small number of protons ($p$) and neutrons ($n$), and DM particles---all in thermal equilibrium with each other. 
The SM particles were initially tightly coupled via electromagnetic and weak interactions, but as the universe expanded and cooled, the rate of weak interactions dropped, resulting in neutrino decoupling at the temperature of $\sim$few MeV \cite{1992NuPhB.374..392E, 2002PhR...370..333D, 2002PhRvD..65h3006H}. Later on, at $T_{\gamma}$$\sim$$m_e$, the rest of the plasma received an energy injection from electron-positron annihilation; within the standard cosmological model, this process heats photons with respect to neutrinos, leading to the present-day neutrino--to--photon temperature ratio of $(T_{\nu}/T_{\gamma})_{0} \simeq (4/11)^{1/3} \simeq 1.4$ \footnote{This result is under the assumption of instantaneous neutrino decoupling.}. Soon after neutrino decoupling, at $\sim$0.7 MeV, the weak interactions could no longer sustain neutron--proton chemical equilibrium; the neutron number density began to plummet due to neutron decay, up until the onset of BBN. Once the universe cooled down enough for the formation of helium--4, deuterium, helium--3, and lithium to take place, the leftover free neutrons were captured (and preserved) within light nuclei. The time elapsed until the universe expanded and cooled enough for nuclei to form is thus the main parameter determining $Y_\mathrm{p}$ and $Y_\mathrm{D}$. 

%%%%%%%%%%%%
\begin{figure}[h]
\begin{center}
\includegraphics[width=0.45\textwidth]{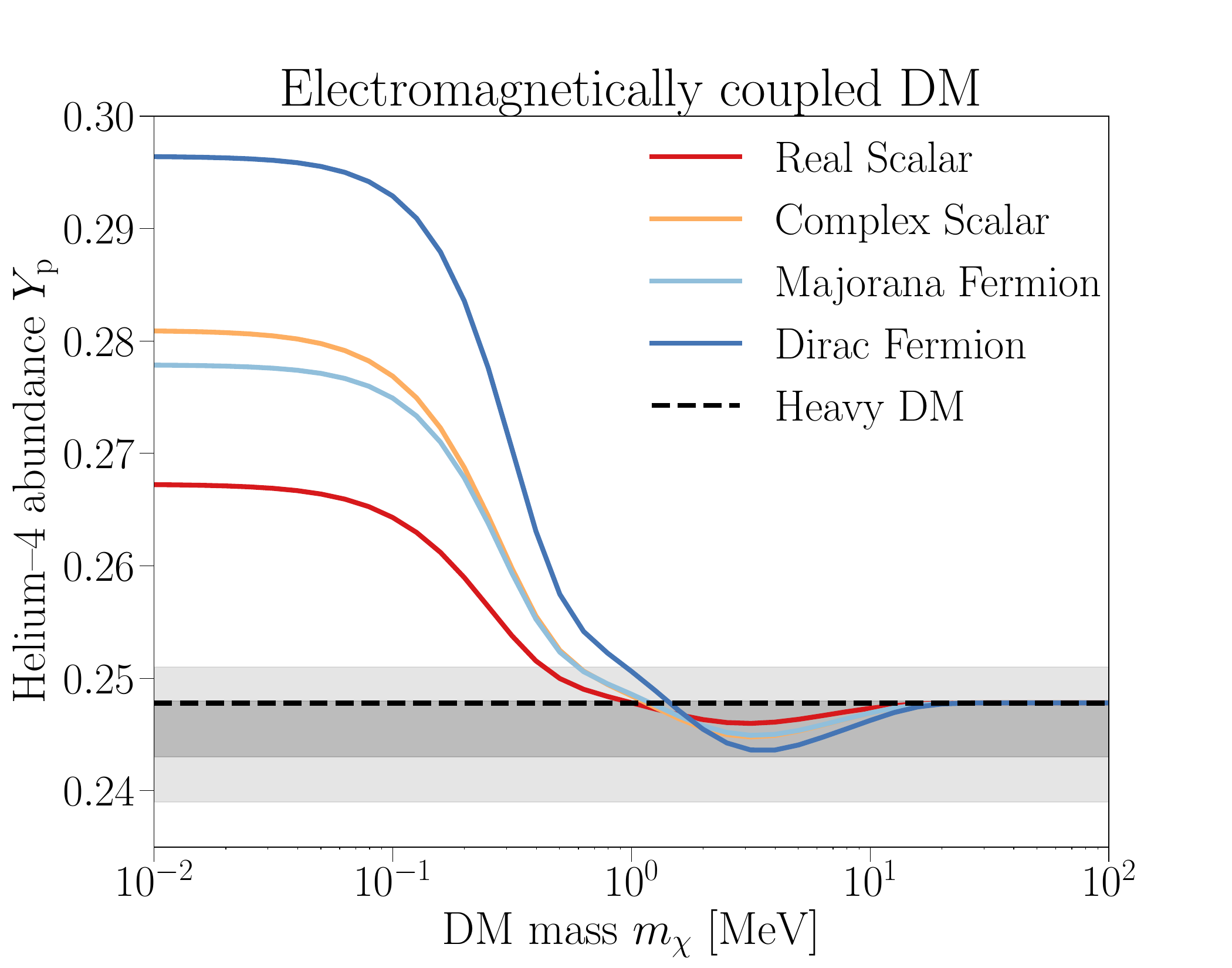}\\
\includegraphics[width=0.45\textwidth]{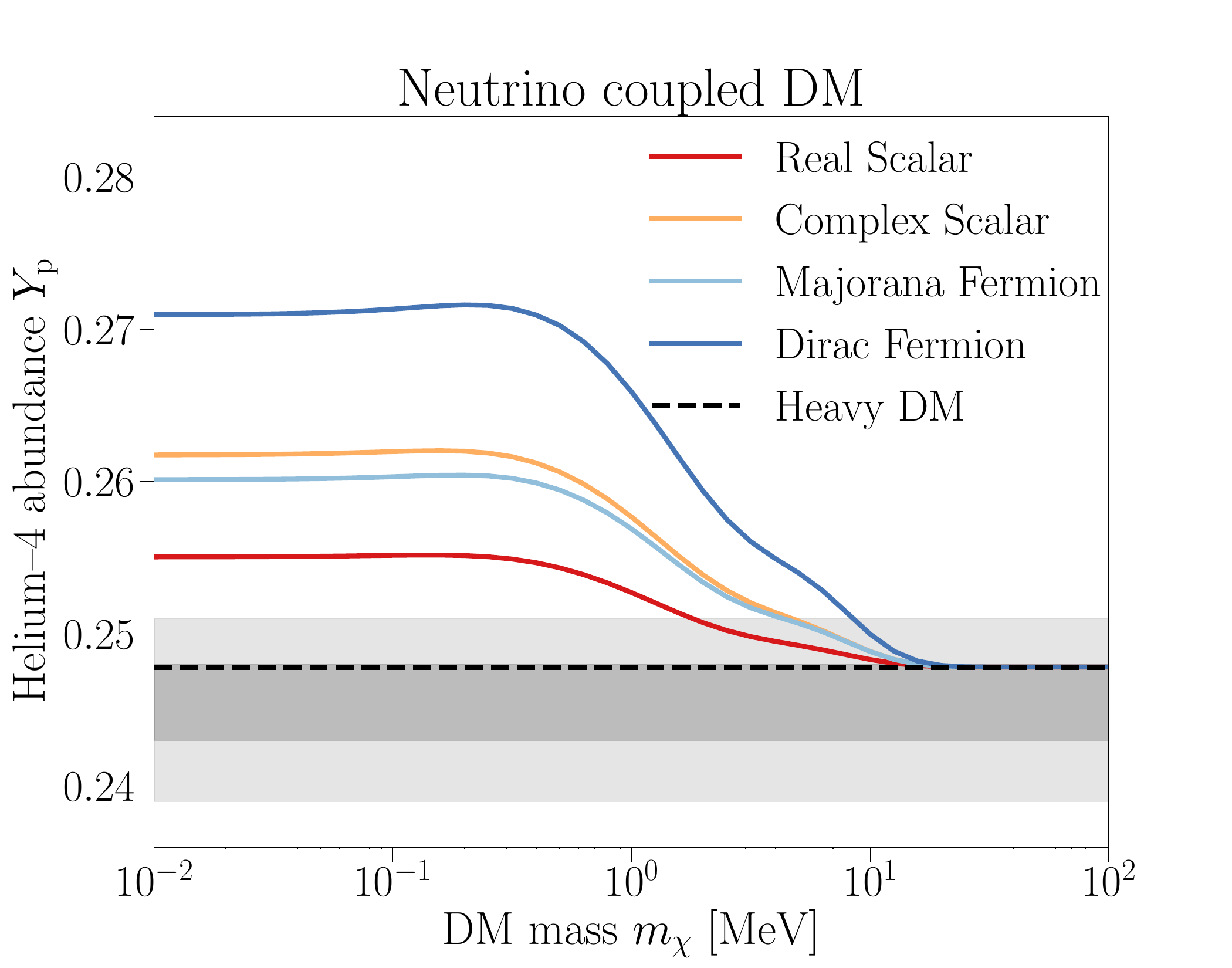}
\end{center}
\caption{Primordial abundance of helium-4, as a function of DM mass $m_{\chi}$, obtained using \texttt{AlterBBN} code. We assume only the standard neutrino species ($\Delta N_{\nu}=0$) and use $\Omega_b h^2=0.0224$ \cite{2020A&A...641A...6P}. The abundances are shown for two sets of models: with DM coupling electromagnetically (top panel) and to the SM neutrinos (bottom panel). The dashed black line is for the standard BBN scenario, where DM has a large mass $m_{\chi}\gtrsim20$ MeV. The grey bands show the current measurement uncertainty on $Y_\mathrm{p}$ from %inferred from spectra of high--redshift absorption systems in 
Ref.~\cite{2020PTEP.2020h3C01P}.}\label{fig:Yp}
\end{figure}
%%%%%%%%%%%%
%%%%%%%%%%%%
\begin{figure}[h]
\includegraphics[width=0.45\textwidth]{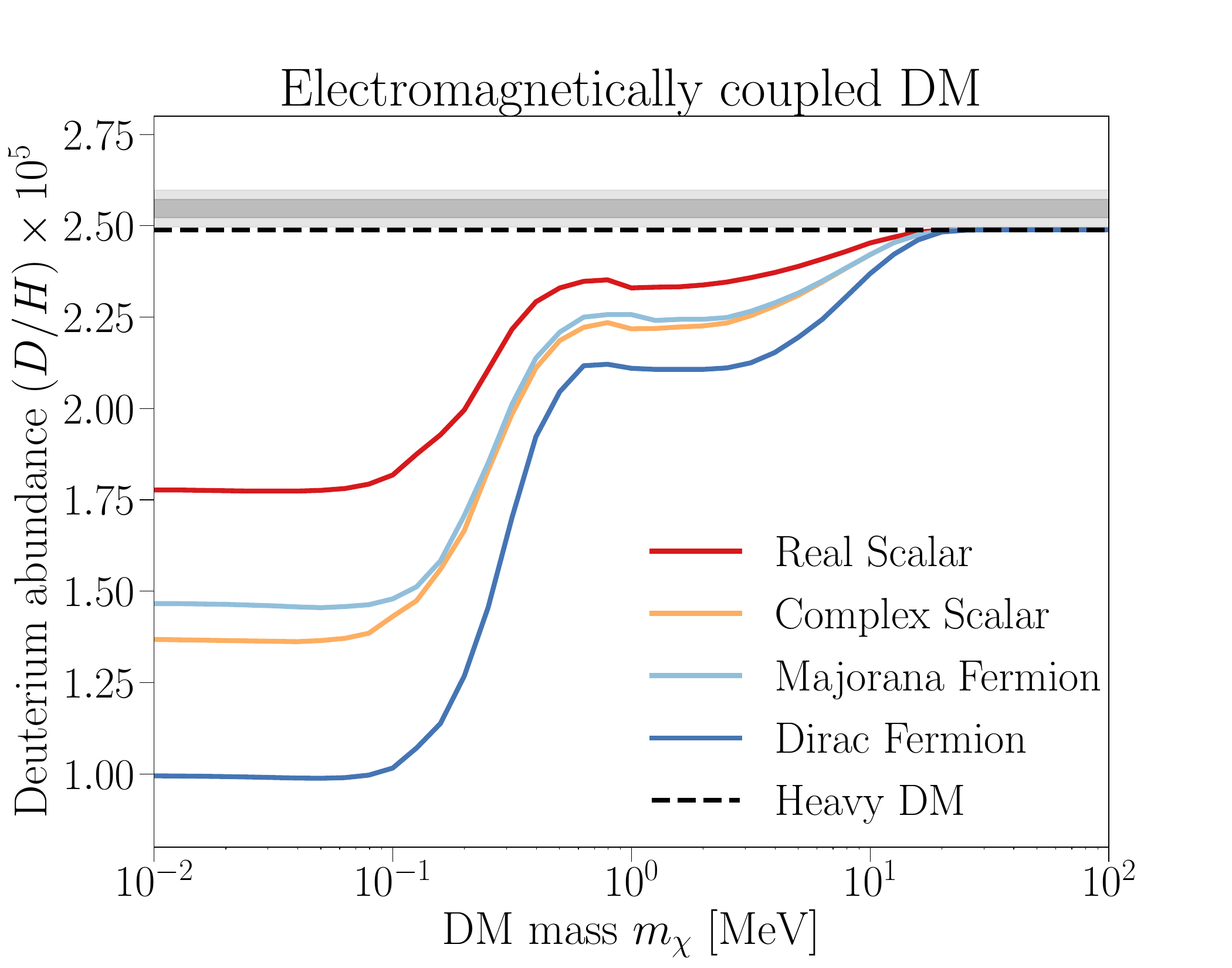}\\
\includegraphics[width=0.45\textwidth]{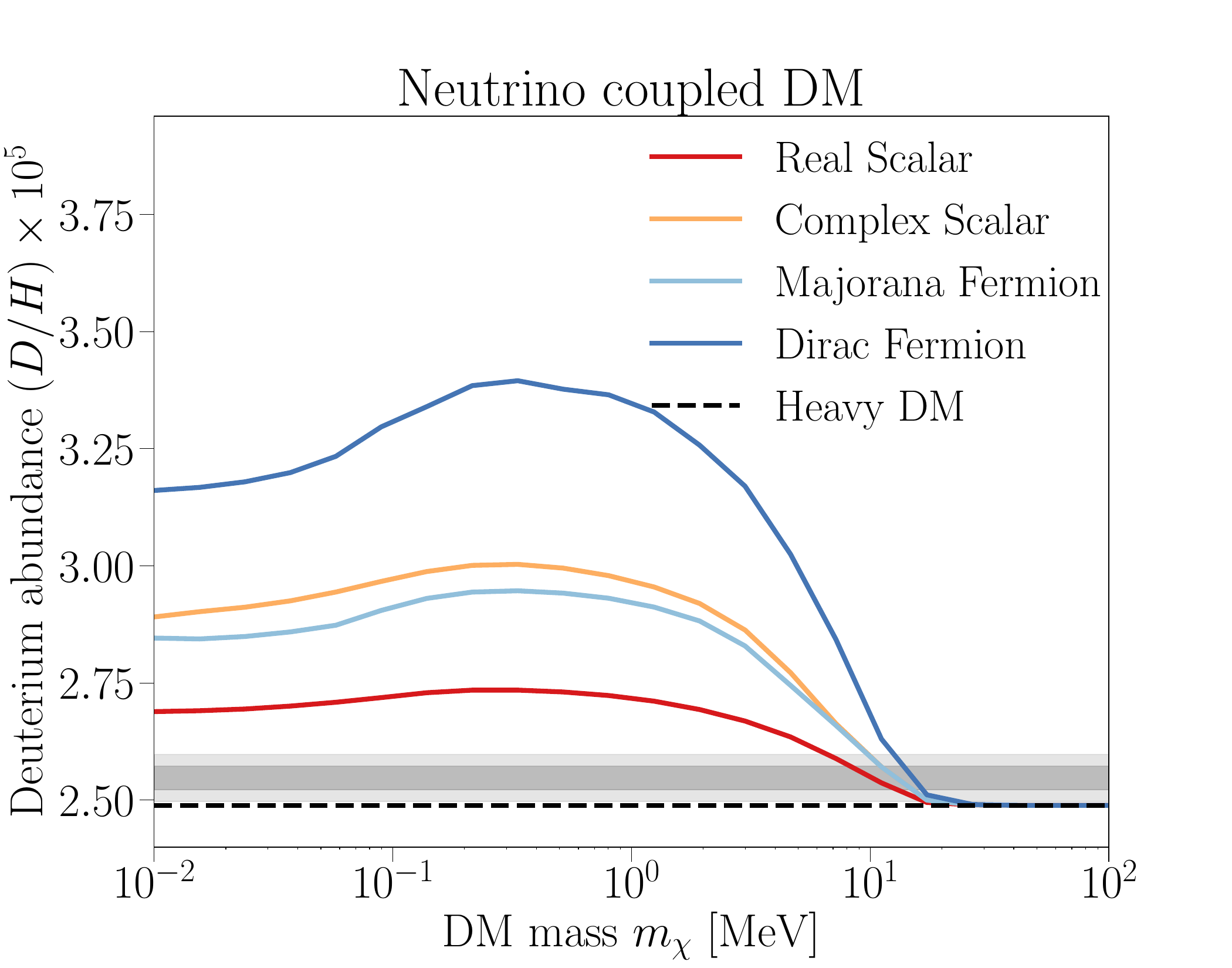}
\caption{Same as Fig.~\ref{fig:Yp}, except for primordial abundance of deuterium $Y_\mathrm{D}=(D/H) \times 10^5$.}\label{fig:D}
\end{figure}
%%%%%%%%%%%%

The rate of expansion is in turn controlled by the content of the universe, including the amount of DM. If DM is heavy enough, it is non-relativistic during BBN and thus contributes negligibly to the overall energy density at the time of BBN. However, if DM is as light as $\sim$20 MeV or less, it may transition to being non-relativistic during BBN, affecting this process. In this case, DM can no longer be efficiently produced within the thermal bath during BBN, and its annihilation into other species takes place, transferring entropy into the rest of the plasma. Depending on the value of $m_\chi$, spin statistic of the specific DM species, and whether it couples to photons, electron--positrons, or neutrinos (the only other radiation species around at the time), the entropy transfer can alter the rate of expansion (and cooling) by different amount, compared to the standard cosmological scenario. Regardless of the modeling specifics, one of the main effects of the presence of light thermal--relic particles is to modify the time at which proton--to--neutron conversion and various other nuclear processes freeze out---and therefore change the values of $Y_\mathrm{p}$ and $Y_\mathrm{D}$ as compared to the standard BBN (SBBN); for a detailed review of the BBN calculation, see Ref.~\cite{Jenssen2016NewAA}. 

In the presence of light thermal--relic DM, the primordial abundance of light elements, including $Y_\mathrm{p}$ and $Y_\mathrm{D}$, depends on three cosmological parameters: baryon--to--photon number density ratio $\eta$ (or, equivalently, the present-day baryon energy density $\Omega_b h^2$), the effective number of new relativistic species beyond the SM neutrinos, $\Delta N_{\nu}$ (where $\Delta N_{\nu}=0$ in SBBN), and DM mass $m_{\chi}$. To quantify the impact of DM mass on the synthesis of the light elements, we use the publicly-available \texttt{AlterBBN} code \cite{2012CoPhC.183.1822A, 2018arXiv180611095A}. \texttt{AlterBBN} enables high-accuracy  predictions for the abundance of light elements generated during BBN in different cosmological scenarios, including the presence of light DM particles\footnote{Apart from AlterBBN, there are other public BBN codes, such as the Fortran program PArthENoPE \cite{2008CoPhC.178..956P, 2018CoPhC.233..237C} and the Mathematica code PRIMAT \cite{2018PhR...754....1P}. We have verified that the specific choice of BBN code does not significantly affect our reported results.}.
In Figs.~\ref{fig:Yp} and \ref{fig:D}, we show the predicted primordial abundances in cosmologies with light thermally-coupled DM, as a function of $m_\chi$, for the two DM scenarios considered in this work, obtained using \texttt{AlterBBN}; these Figures also show the current best bounds on these quantities, obtained from Lyman-$\alpha$ forest measurements and presented in Ref.~\cite{2020PTEP.2020h3C01P}. 

From Fig.~\ref{fig:Yp}, we see that DM with $m_{\chi}\gtrsim 20$ MeV has no significant effect on $Y_\mathrm{p}$, because it annihilates before the weak--interaction freezeout. In contrast, lower--mass DM annihilates after neutrino decoupling and can significantly alter the BBN yields. Beyond this global trend, there are additional subtleties captured in the BBN calculation shown here. For example, in the electromagnetically coupled DM scenario, the expansion rate can slow down for a narrow range of masses, increasing the time available for neutron decay, but also leading to a slower conversion rate between neutrons and protons, due to the relative decrease in neutrino energy density, as compared to photons \cite{1986PhRvD..34.2197K,2014PhRvD..89h3508N}. These two effects nearly cancel out, leading to a very slightly reduced relic abundance in the range $2\text{ MeV}\lesssim m_{\chi}\lesssim 20$ MeV, as compared to SBBN (as shown in the top panel of Fig.~\ref{fig:Yp}). Lighter DM particles simply contribute to the expansion rate and result in a higher value of $Y_\mathrm{p}$. In the neutrino coupled DM scenario, neutrino--to--photon temperature ratio is increased compared to SBBN, speeding up the rate of weak interactions, and driving up the value of $Y_\mathrm{p}$ at all masses below $\sim$20 MeV \cite{2015PhRvD..91h3505N}. 

From Fig.~\ref{fig:D}, we see that $Y_\mathrm{D}$ is altered relative to SBBN only for light DM with $m_{\chi}\lesssim 20$ MeV. The electromagnetically coupled DM particles reduce the neutrino--to--photon temperature ratio, resulting in slower expansion rate at a fixed temperature $T_{\gamma}$, leaving more time for deuterium destruction, and reducing its relic abundance. Conversely, the increased neutrino--to--photon temperature ratio in the neutrino coupled DM scenario leads to a higher value of $Y_\mathrm{D}$.

%%%%%%%%%%%%%%%%%%%%%%%%%%%%%%%%%%%%%%%
\section{Effects on CMB} \label{sec:CMB}

\begin{figure}
\includegraphics[width=0.48\textwidth]{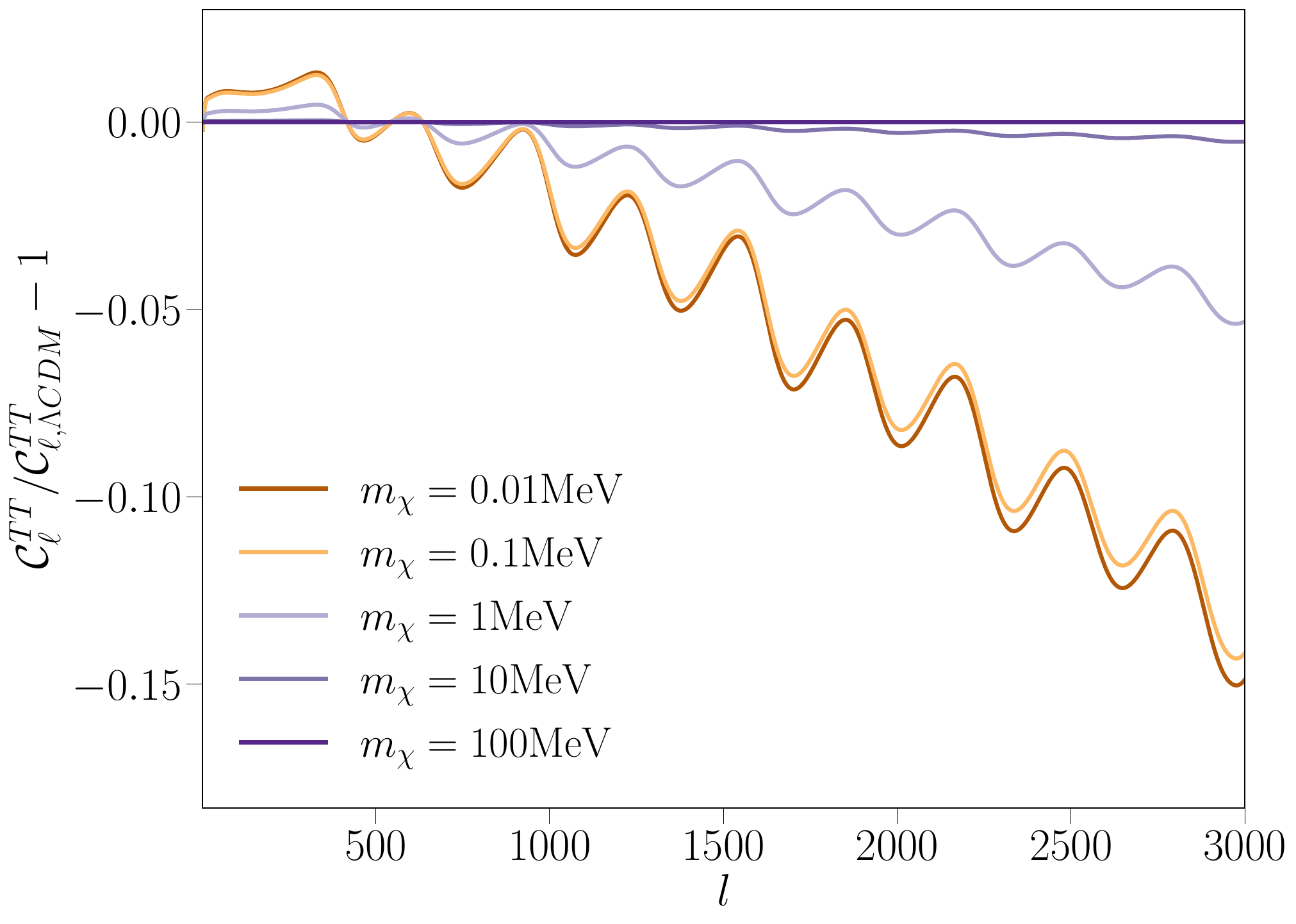}
\caption{The ratio of the CMB temperature power spectrum for a cosmology featuring light electromagnetically coupled Dirac Fermion DM, with a given mass $m_\chi$, as compared to the $\Lambda$CDM scenario (with all other parameters kept at their best-fit \textit{Planck} values). Suppression of power at small angular scales occurs due to an increased primordial yield of helium in light--DM cosmologies. Smaller particle masses lead to a more prominent suppression, as they alter the expansion rate during BBN more significantly. For these purposes, masses above $\gtrsim20$ MeV behave as standard (heavy) CDM.}\label{fig:cl}
\end{figure}

Light thermal--relic particles can leave imprints on the CMB through two distinct effects. First, as discussed in Sec.~\ref{sec:BBN}, they can alter the process of BBN and change $Y_\mathrm{p}$, in turn affecting the temperature and polarization anisitropy. In particular, since helium--4 recombines at a lower temperature than hydrogen, increasing $Y_\mathrm{p}$ (and fixing all other cosmological parameters) reduces the number of free electrons present in the universe at a particular fixed temperature. Thus, the mean free path of photons increases, resulting in increased diffusion length and damping of the small--scale anisotropy in the CMB \cite{2004PhRvD..69b3509T}. As a result, enhancing $Y_\mathrm{p}$ due to the presence of light thermal--relic DM during BBN generically leads to a suppression of the CMB power spectra. We modify the standard Boltzmann code \texttt{CLASS} \cite{2011arXiv1104.2932L, class2011} in order to include the effects of $m_\chi$, as described in more detail below; using the output of our modified version of \texttt{CLASS}, we illustrate the effect on the temperature anisotropy in Fig.~\ref{fig:cl}, for a fixed total $N_\mathrm{eff}=3.046$.
%%%%%%%%%%%%
%%%%%%%%%%%%

In addition to the effect on $Y_\mathrm{p}$, light thermal relics can affect the overall budget of radiation, quantified by the $N_\mathrm{eff}$ parameter, which controls the expansion rate at later epochs and affects the CMB anisotropy. 
For example, DM annihilating into photons \textit{after} neutrino decoupling can heat photons relative to the decoupled neutrinos, reducing the present--day ratio of neutrino--to--photon temperature (assuming that the present--day temperature of the CMB is kept fixed), corresponding to a reduced value of $N_\mathrm{eff}$ \cite{2013PhRvD..87j3517S}. Conversely, DM annihilating to neutrinos heats up neutrinos relative to photons, increasing $N_\mathrm{eff}$. 

%%%%%%%%%%%%
\begin{figure}
\includegraphics[width=0.45\textwidth]{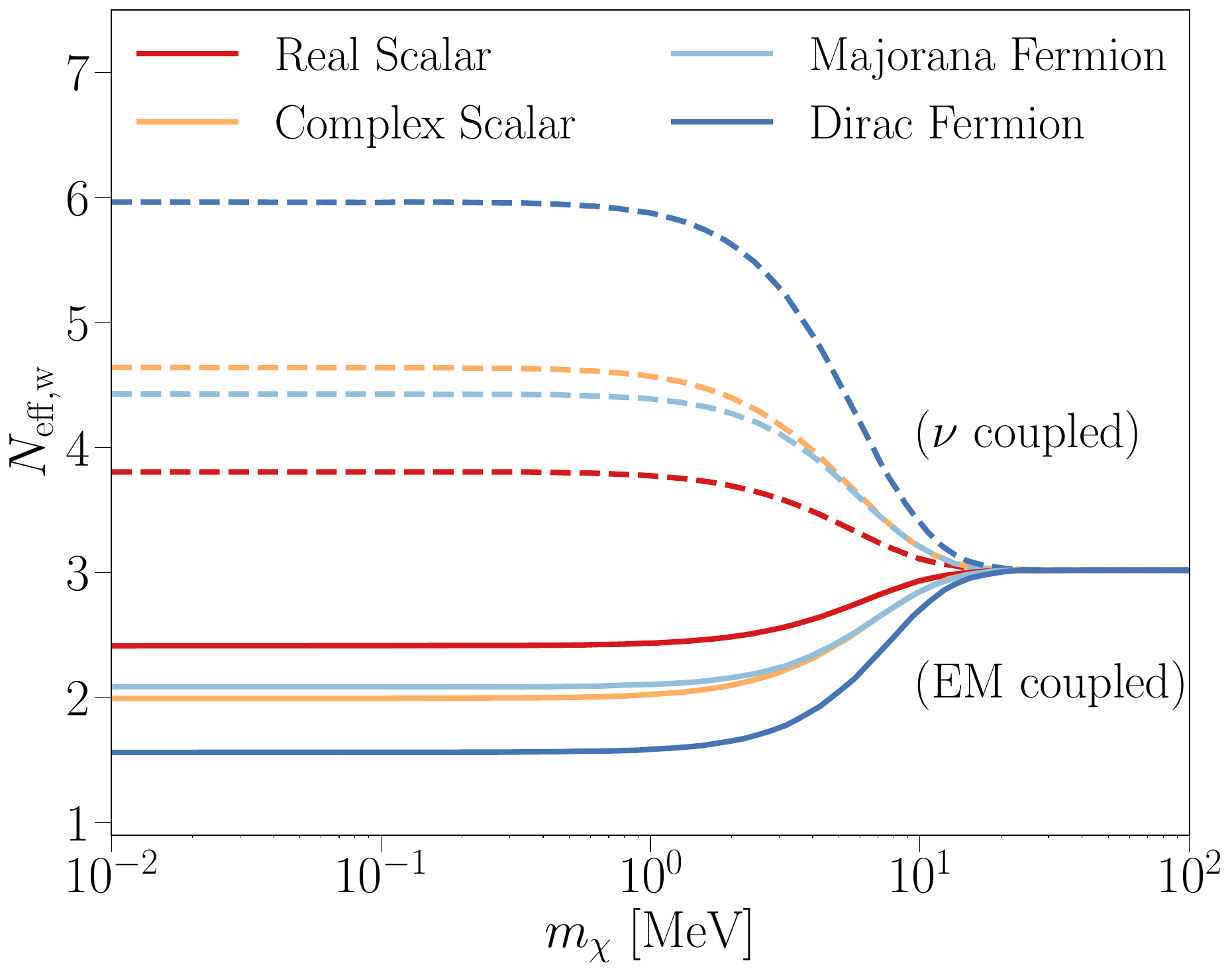}
\caption{The number of relativistic degrees of freedom $N_\mathrm{eff,w}$ in a cosmology featuring light thermal--relic DM, as a function of $m_{\chi}$ in the electromagnetically coupled (dashed lines) and neutrino coupled (solid lines) scenarios. Since neutrino coupled DM particles increase the temperature of neutrinos relative to photons, they also increase the effective number of relativistic species, while the opposite is true for DM that annihilates into photons. The curves are obtained using \texttt{AlterBBN} code, assuming only the standard three neutrino species are present.}\label{fig:Neff}
\end{figure}
%%%%%%%%%%%%
In standard cosmology, with only the thee SM neutrino species and SBBN, $N_\mathrm{eff}=3.046$.
On the other hand, in light--DM cosmologies, we can define 
(following Ref.~\cite{2013PhRvD..87j3517S})
\begin{equation}
    N_\mathrm{eff}(m_\chi,\Delta N_\nu)\equiv N_\mathrm{eff, w}(m_\chi)(1+\Delta N_{\nu}/3)
    \label{eq:Neff}
\end{equation} 
to be the total number of relativistic degrees of freedom, where  
\begin{equation}
    N_\mathrm{eff,w}\equiv 3\left[\frac{11}{4}\left(\frac{T_\nu}{T_\gamma}\right)_0^3\right]^\frac{4}{3}
\end{equation}
includes the contribution from the SM neutrinos, as well as contribution from light DM, and the subscript zero denotes the present time.
$\Delta N_{\nu}$ is the effective number of any \textit{additional} new (neutrino--like) relativistic species that may be present in the universe. We note that $N_\mathrm{eff,w}$ depends on the DM mass $m_{\chi}$, but differs for different particle models, as illustrated in Fig.~\ref{fig:Neff}, using \texttt{AlterBBN} code (for $\Delta N_{\nu}=0$).  We see that particles with $m_{\chi}\gtrsim 20$ MeV have no significant effect on $N_\mathrm{eff,w}$, because their annihilation is complete prior to the decoupling of the SM neutrinos, and therefore does not affect the neutrino--to--photon temperature, giving $N_\mathrm{eff, w}=3.046$. Interestingly, presence of lighter DM particles coupled to photons can conceal a possible existence of additional relativistic species, by offsetting their effect on $\Delta N_{\nu}$; we consider this possibility in Sec.~\ref{sec:constraints}. 

In order to account for the effects of $m_\chi$ in cosmologies with thermal--relic DM, both through its effects on the BBN yields and the effects on $N_\mathrm{eff}$, we modify \texttt{CLASS} code \cite{2011arXiv1104.2932L,class2011}. In the standard version of this code, the value of $Y_\mathrm{p}$ assumes a standard BBN scenario, where $Y_\mathrm{p}$ depends only on $\Omega_b h^2$ and $\Delta N_{\nu}$ (for a fixed present-day CMB temperature). In the modified version of \texttt{CLASS}, we use an updated prediction for $Y_\mathrm{p}$, as a function of $\Omega_b h^2$, $\Delta N_{\nu}$, and $m_\chi$, obtained from \texttt{AlterBBN}. Fig.~\ref{fig:cl} illustrates the output of our modified \texttt{CLASS} code, including both effects discussed in this Section (note that we fix the effective number of neutrino at $N_\mathrm{eff}=3.046$ in this Figure). 

%%%%%%%%%%%%%%%%%%%%%%%%%%%%%%%%%%%%%%%
\section{Current constraints}\label{sec:constraints}

We consider the most recent CMB measurements from ACT, SPT, and \textit{Planck}, as well measurements of primordial element abundances from spectroscopic measurements of high--redshift absorption systems, in order to infer $m_\chi$. We now describe the data sets, our analysis method, and the constraints we obtain.

%%%%%%%%%%%%%%%%%%%%%%%%%%%%%%%%%%%%%%%
\subsection{Data}

\textbf{ACT DR4}: We use the multifrequency TT, EE, and TE power spectra from ACT Data Release 4 (DR4) \cite{2020JCAP...12..047A, 2020JCAP...12..045C}, implemented within the \texttt{actpollite\_dr4} likelihood \footnote{\url{https://github.com/ACTCollaboration/pyactlike}}. These data products are derived from a four--year survey, with the power spectrum measurements reconstructed from the deepest 5400 $\text{deg}^2$ of the sky. This data provide high resolution measurements of the polarization anisotropy, complementing the data from \textit{Planck}. In the ACT likelihood, the covariance of the foreground--marginalized CMB power spectra already includes the effects of noise, foreground uncertainty, beam and calibration uncertainties, with one nuisance parameter $y_{p}$ included to marginalize over an overall polarization efficiency; we allow this variable to vary in a range centred at 1. ACT alone cannot constrain the optical depth to reionization $\tau$, as it is mainly determined by low--$\ell$ polarization power spectra; thus, when analyzing ACT data alone, we assume a Gaussian prior on $\tau=0.065\pm 0.015$, following Ref.~\cite{2020JCAP...12..047A}.

\textbf{SPT--3G}: We use the most recent publicly--available measurements of EE and TE power spectra from the SPT--3G survey \cite{2021PhRvD.104b2003D}. These power spectra are obtained from the observations of a $1500\text{deg}^2$ region of the sky, taken over four months (in 2018) at three frequency bands centered on 95, 150, and 220 GHz. The SPT likelihood is publicly available\footnote{\url{https://pole.uchicago.edu/public/data/dutcher21}} and includes the effects of the aberration due to relative motion with respect to the CMB rest frame \cite{2014PhRvD..89b3003J}, super-s-ample lensing \cite{2014PhRvD..90b3003M}, polarized foregrounds, uncertainty in the calibration of the bandpowers, and uncertainty in the beam measurements. The priors on many of these terms are listed in the Table III of Ref.~\cite{2021PhRvD.104b2003D}. Similar to ACT, SPT has the advantage of high resolution and high sensitivity specifically in polarization measurements on small scales. Since the low--$\ell$ polarization are not probed by this experiment, we adopt a \textit{Planck}--based Gaussian prior of $\tau=0.0543\pm 0.007$\footnote{Widening the prior to $\tau=0.065\pm 0.015$ has no significant effect on cosmological parameter constraints} \cite{2020A&A...641A...6P} when analyzing SPT data alone, following Ref.~ \cite{2021PhRvD.104b2003D}.

\textbf{\textit{Planck} 2018}: We use both low--$\ell$ and high--$\ell$ multifrequency power spectra TT, TE, and EE from \textit{Planck} PR3 (2018) \cite{2020A&A...641A...6P}, available through the Legacy Archive for Microwave Background Data Analysis (LAMBDA\footnote{\url{https://lambda.gsfc.nasa.gov/}}). We rely only on the \textit{lite} (pre-marginalized) likelihood for high-$\ell$ TTTEEE, for computational efficiency; using the full likelihood does not lead to appreciable changes in the inferred parameter values for models we consider here \cite{2020A&A...641A...6P}. \textit{Planck} data provides lower noise level at large scales, as compared to the ground-based observations, enabling complementary constraints on cosmological parameters.

\textbf{\textit{Planck}+ACT+SPT}: When combining ACT DR4 and SPT--3G with \textit{Planck} 2018 data, the covariance between SPT and \textit{Planck} is expected to be negligible since the SPT observation region is only a small fraction of the Planck field. Following Ref.~\cite{2020JCAP...12..047A}, we minimize covariance between ACT and \textit{Planck} by discarding large scales ($\ell<1800$) in TT (and with no cuts in TE and EE). The correlations between SPT and ACT can be ignored because they observe different parts of the sky. Note that we rely on \textit{Planck} low-$\ell$ EE to constrain $\tau$ in this case.

\textbf{Primordial element abundances}: In addition to the CMB anisotropy, we also consider the measurements of the primordial helium--4 and deuterium abundances, derived from Lyman--$\alpha$ forest observations and reported in Ref.~\cite{2020PTEP.2020h3C01P}, where $Y_\mathrm{p}=0.245\pm{0.003}$ at 68\% CL, and the bound on deuterium is $Y_\mathrm{D}=2.547\pm{0.025}$. 
Besides the observational uncertainties, we also account for theoretical uncertainty related to uncertainties on various nuclear reaction rates, $\sigma_{Y_\mathrm{p}^{\mathrm{th}}}=1.6\times 10^{-4}$ and $\sigma_{Y_\mathrm{D}^{\mathrm{th}}}=0.03$ \cite{Jenssen2016NewAA,2018arXiv180611095A}, as discussed in the following.

%%%%%%%%%%%%%%%%%%%%%%%%%%%%%%%%%%%%%%%
\subsection{Method}

The first step in our analysis is to determine the values of $N_\mathrm{eff}$ and $Y_\mathrm{p}$ that are consistent with CMB data, without reference to $m_\chi$. To do this, we carry out a series of Markov Chain Monte Carlo (MCMC) runs within the \texttt{Cobaya} sampling framework \cite{2021JCAP...05..057T, 2019ascl.soft10019T}, individually and jointly applied to ACT, SPT, and \textit{Planck} data. We utilize \texttt{mcmc} sampler, and employ the convergence criterion $R-1=0.01$, where R is the Gelman--Rubin threshold \cite{1992StaSc...7..457G}. 
In each MCMC run, we sample the posterior distributions of the six standard cosmological parameters (baryon density $\Omega_bh^2$, DM density $\Omega_ch^2$, acoustic scale $100\theta$, reionization optical depth $\tau$, scalar spectral index $n_s$, and amplitude of the scalar perturbations $A_s$), with the addition of $Y_\mathrm{p}$ and $N_\mathrm{eff}$. For each parameter, we employ broad priors listed in Table \ref{tab:prior}. Importantly, we allow $Y_\mathrm{p}$ to vary in an unconstrained way, in order to determine values that are consistent with CMB observations, regardless of BBN predictions. Results of the MCMC runs are shown as 68\% and 95\% CL contours in Fig.~\ref{fig:YN_0}, and are consistent with previous analyses \cite{2020A&A...641A...6P,2020JCAP...12..047A,2021PhRvD.104h3509B}.
%%%%%%%
\begin{table}
\caption{Priors on cosmological parameters used in the likelihood analysis.\label{tab:prior}}
\setlength{\tabcolsep}{8mm}{
\begin{tabular}{|c|c|}
\hline
Parameter & Prior \\
\hline
$\Omega_bh^2$ & [0.005, 0.1]\\
\hline
$\Omega_ch^2$ & [0.001, 0.99]\\
\hline
$100\theta$ & [0.5, 10]\\
\hline
$\tau$ \footnote{For ACT DR4 data alone, we use a Gaussian prior of $\tau=0.065\pm 0.015$. For SPT--3G data alone, we use a Gaussian prior of $\tau=0.0543\pm 0.007$.} & [0.01, 0.8]\\
\hline
$n_s$ & [0.8, 1.2]\\
\hline
$\text{log}(10^{10}A_s)$ & [1.61, 3.91]\\
\hline
%$m_{\chi}$ [MeV] & [0.01, 100]\\
$Y_\mathrm{p}$ & [0.1, 0.5]\\
\hline
$N_\mathrm{eff}$ & [1, 5]\\
\hline
\end{tabular}}
\end{table}
%%%%%%%

Once we obtain the full posterior probability distributions using CMB data, the second step we take is to relate the preferred values of $\Omega_bh^2$, $N_\mathrm{eff}$, and $Y_\mathrm{p}$ to the corresponding allowed values of $m_\chi$ and $\Delta N_\nu$ and obtain their bounds under each DM model. To do this, we approximate our posterior distributions as Gaussian and employ the following chi--squared statistic to determine the relevant CL intervals
\begin{equation}\label{eq:chiCMB}
\chi_{\rm{CMB}}^2  = \left(\textbf{X} - \textbf{X}_{\rm{obs}}\right) \textbf{Cov}^{-1} \left(\textbf{X} - \textbf{X}_{\rm{obs}}\right)^T, 
\end{equation}
where $\textbf{X}=\{\Omega_b h^2, N_\mathrm{eff}, Y_\mathrm{p}\}$ is a vector of three relevant cosmological parameters; $\textbf{X}_{\rm{obs}}$ is the inferred maximum--likelihood value of the three parameters; and $\textbf{Cov}$ is the covariance matrix in the 3d parameter space, derived from the sample chains. Using \texttt{AlterBBN}, we recover the mapping $N_\mathrm{eff}(m_\chi, \Delta N_\nu)$ and $Y_\mathrm{p}(m_\chi, \Delta N_\nu, \Omega_b h^2)$ for each DM model of interest. Finally, using this mapping, we identify the appropriate values of $\chi_{\rm{CMB}}^2$ with 95\% CL intervals for $m_\chi$ and $\Delta N_\nu$.

To add the primordial abundance measurements from Lyman--$\alpha$ forest spectra requires additionally imposing the constraints on $Y_\mathrm{p}$ and $Y_\mathrm{D}$. To do this, we compute the primordial--abundance chi--squared statistic $\chi_{\rm{PA}}^2$, as
\begin{equation}\label{eq:chiBBN}
\chi_{\rm{PA}}^2 = \frac{\left[Y_\mathrm{p} - Y_\mathrm{p}^{\rm obs}\right]^2}{{\sigma_{Y_\mathrm{p}^{\rm th}}}^2 + {\sigma_{Y_\mathrm{p}^{\rm obs}}}^2}+\frac{\left[Y_\mathrm{D} - Y_\mathrm{D}^{\rm obs}\right]^2}{{\sigma_{Y_\mathrm{D}^{\rm th}}}^2 + {\sigma_{Y_\mathrm{D}^{\rm obs}}}^2},
\end{equation}
which includes both observational and theoretical uncertainties. To infer combined bounds from CMB and primordial--abundance measurements, we evaluate the sum of the relevant statistics, $\chi_{\rm{tot}}^2\equiv \chi_{\rm{CMB}}^2 + \chi_{\rm{PA}}^2$.

%%%%%%%%%%%%%%%%%%%%%%%%%%%%%%%%%%%%%%%
\subsection{CMB--only bounds}\label{sec:cmb_bounds}

%%%%%%%%%%%%%%
\begin{figure}
\includegraphics[width=0.42\textwidth]{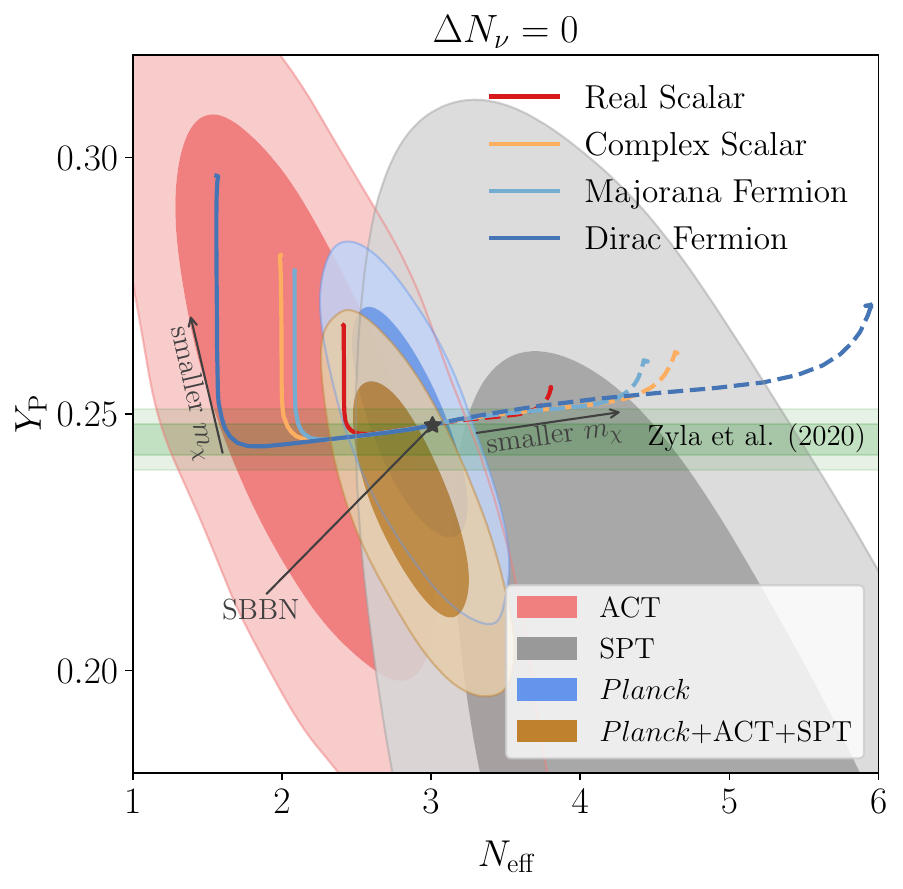}
\caption{95 \% CL bounds are shown as lighter colored areas in the $N_\mathrm{eff}$--$Y_\mathrm{p}$ plane, as derived from 2d marginalized posterior distribution, obtained from different CMB data sets, denoted in the legend. The horizontal band represents the helium--4 abundance measurement from Ref.~\cite{2020PTEP.2020h3C01P}. Theoretical predictions are shown as solid and dashed lines, for the electromagnetically and neutrino coupled DM, respectively, both for $\Delta N_{\nu}=0$ (no other relativistic species are allowed); DM spin statistic is denoted in the legend. All theoretical curves converge to a single point that corresponds to standard BBN with $m_{\chi}\gtrsim20$ MeV. Along each theoretical curve, $m_{\chi}$ varies from the this point to $\lesssim$0.01 MeV at the upper left (solid) and upper right (dashed) end of the lines. }\label{fig:YN_0}
\end{figure}
%%%%%%%%%%%%%%
%%%%%%%%%%%%%%
\begin{figure}
\includegraphics[width=0.42\textwidth]{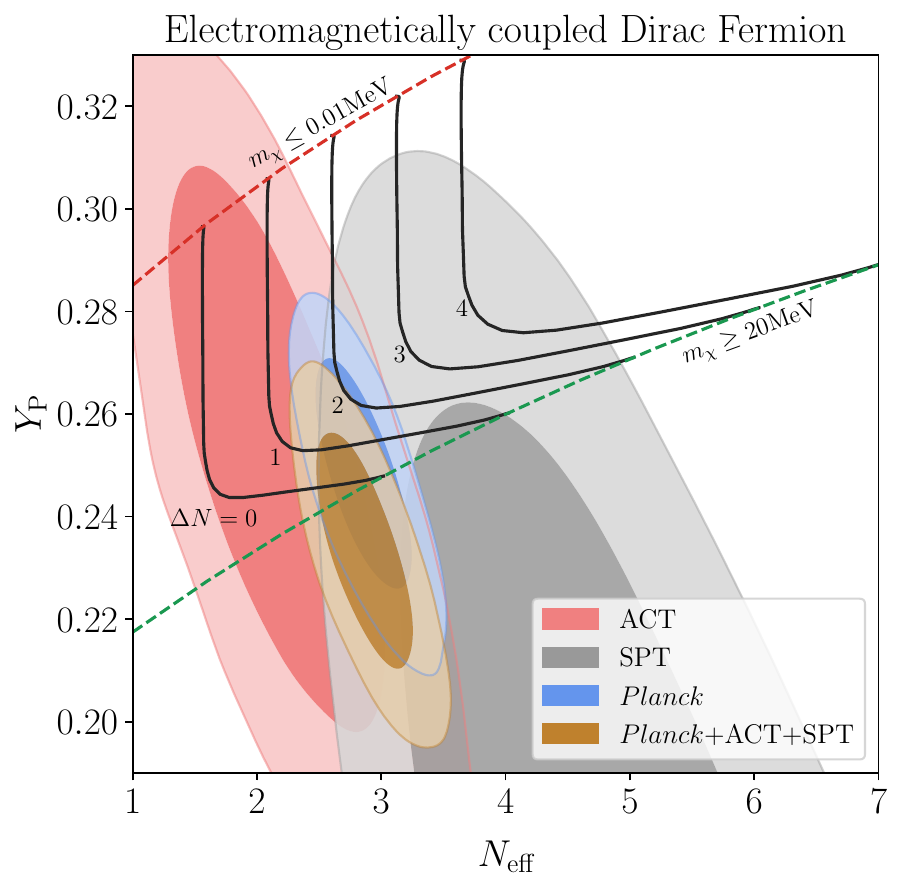}\\
\includegraphics[width=0.42\textwidth]{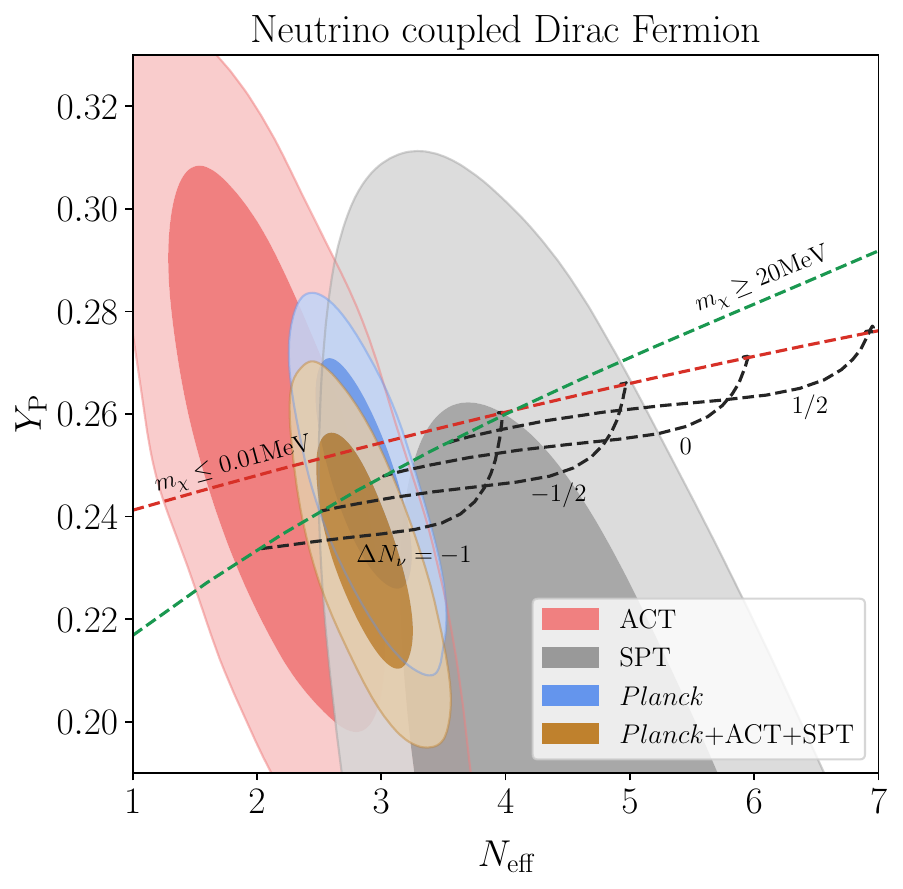}
\caption{The constraints on $N_\mathrm{eff}$ and $Y_\mathrm{p}$ derived from CMB data are the same as in Fig.~\ref{fig:YN_0}. The theoretical curves are shown for the Dirac Fermion DM that couples electromagnetically (top panel) or to neutrinos (bottom panel).
Theoretical predictions are shown as black lines for different values of $\Delta N_{\nu}$. Along each line, $m_{\chi}$ varies from $\gtrsim$20 MeV to $\lesssim$0.01 MeV, as labeled in the plots.}\label{fig:YN_3}
\end{figure}
%%%%%%%%%%%%%%%%

Figs.~\ref{fig:YN_0} and \ref{fig:YN_3} show CMB--only bounds in the $N_\mathrm{eff}$--$Y_\mathrm{p}$ plane, with darker and lighter shaded areas representing 68\% and 95\% CL regions, respectively. The bounds are obtained by varying the full set of eight cosmological parameters shown in Table \ref{tab:prior}. The $N_\mathrm{eff}$ best--fit value is higher for SPT data than for \textit{Planck}, which in turn is higher than that for ACT. 
However, at the 2$\sigma$ level, all three data sets are consistent with each other. 
We note that the bounds in these plots are \textit{not} dependent on DM model, and no DM parameters are fit to the data at this stage; rather, the bounds on $N_\mathrm{eff}$ and $Y_\mathrm{p}$ are used to obtain posterior distribution for $m_\chi$ in the second stage of the analysis. For illustration, the Figures also show theoretical relationship between $m_\chi$, $\Delta N_\nu$, and $Y_\mathrm{p}$; this relationship is illustrated in Figs.~\ref{fig:Yp} and \ref{fig:Neff}, and Eq.~(\ref{eq:Neff}) (for $\Omega_b h^2=0.02237$). 

In Fig.~\ref{fig:YN_0}, we assume no other relativistic degrees of freedom and fix $\Delta N_{\nu}=0$. We then illustrate the theoretical relationship between $N_\mathrm{eff}$ and $Y_\mathrm{p}$ for DM coupling either electromagnetically (solid lines) or to SM neutrinos (dashed lines), while the spin statistic of DM particles is denoted in the legend. Mass $m_{\chi}$ varies along each curve from $\lesssim$0.01 MeV (at the far left and far right ends of the theoretical lines) to $\gtrsim$20 MeV (at the center of the panel, where the dashed and the solid lines converge to a point). In the former, the effects on $N_\mathrm{eff}$ and $Y_\mathrm{p}$ saturate at $\sim$0.01 MeV since lighter particles annihilate \textit{after} the end of BBN, while the latter case represents the SBBN scenario where DM is too massive to affect BBN through either effect. 

In Fig.~\ref{fig:YN_3}, we focus on one example of DM spin statistic (for a Dirac Fermion) and explore the effect of $\Delta N_\nu$. The two panels correspond to electromagnetically coupled (top panel) and neutrino coupled (bottom panel) particles. The black curves show the theoretical relationship between $N_\mathrm{eff}$ and $Y_\mathrm{p}$ for several values of $\Delta N_{\nu}$, with $m_{\chi}$ varying along each curve from $\lesssim$0.01 MeV (at the red dashed line) to $\gtrsim$20 MeV (at the green dashed line).  We note that any new relativistic degrees of freedom can only contribute positive values of $\Delta N_\nu$, while negative values are included here for illustration purposes only, and do \textit{not} correspond to known physical models. 

In Figs.~\ref{fig:YN_0} and \ref{fig:YN_3}, the allowed values of $m_\chi$ correspond to the parameter space where the theoretical curve (for a given model and a given choice of $\Delta N_\nu$) is consistent with the shaded region of the parameter space, preferred by the CMB data; the values of $m_\chi$ outside these regions are excluded by our CMB analyses. Precise bounds on $m_\chi$ obtained using our $\chi^2$ procedure are listed in Table~\ref{tab:CMB}, both for the case where $\Delta N_\nu=0$ is fixed, and for the case where $\Delta N_\nu$ is varied as a free parameter (and marginalized over).
%%%%%%%%%%%%%%
\begin{table*}
\caption{95\% CL lower bounds on DM mass $m_{\chi}$ (in units of MeV), inferred from \textbf{CMB data only}, both in the absence of other relativistic species ($\Delta N_{\nu}=0$) and allowing for their presence ($\Delta N_{\nu}\geq 0$, where $\Delta N_{\nu}$ is marginalized over). `$-$' indicates that there are no lower bounds. \label{tab:CMB}}
\setlength{\tabcolsep}{2.5mm}{
\begin{tabular}{|l|c|c|c|c|c|c|c|c|}
\hline
\multicolumn{1}{|c|}{} & \multicolumn{4}{c|}{$\Delta N_{\nu}=0$} & \multicolumn{4}{c|}{$\Delta N_{\nu}\geq 0$} \\
\hline
 & ACT & SPT & \textit{Planck} & \textit{Planck}+ACT+SPT & ACT & SPT &\textit{Planck} & \textit{Planck}+ACT+SPT \\
\hline
\multicolumn{9}{|c|}{Electromagnetically coupled DM} \\
\hline
Real scalar & - & 1.94 & - & -  & - & - & - & -  \\
Complex scalar & - & 4.99 & 4.96 & 3.61 & - & -  & 0.06 & 0.21 \\
Majorana Fermion & - & 4.91 & 4.85 & 3.35 & - & - & - & 0.18  \\
Dirac Fermion & - & 7.72 & 7.65 & 6.45  & - & 0.21 & 0.30 & 0.45 \\
\hline
\multicolumn{9}{|c|}{Neutrino coupled DM} \\
\hline
Real scalar & 6.69 & -  & 6.57  &  11.26 & 6.92 & -  &  6.72 & 11.38  \\
Complex scalar & 9.24 & -  & 8.81 & 13.36  & 9.49 & -  & 9.14 & 13.54 \\
Majorana Fermion & 9.21 & -  & 8.78 & 13.29 & 9.42 & - & 9.06 & 13.43  \\
Dirac Fermion & 11.53 & 1.17 &  11.02 & 15.65  & 11.72 &  1.36  & 11.39 & 15.94 \\
\hline
\end{tabular}}
\end{table*}
%%%%%%%%%%%%%%
%%%%%%%%
\begin{table*}
\caption{Same as Table \ref{tab:CMB}, except for either primordial abundance analysis only (denoted as $Y_\mathrm{p}+Y_\mathrm{D}$), or a joint analysis of all CMB data (\textit{Planck}+ACT+SPT) and the primordial abundance measurements. \label{tab:CMB_Yp_YD}}
\setlength{\tabcolsep}{2.5mm}{
\begin{tabular}{|l|c|c|c|c|c|c|}
\hline
\multicolumn{1}{|c|}{} & \multicolumn{3}{c|}{$\Delta N_{\nu}=0$} & \multicolumn{3}{c|}{$\Delta N_{\nu}\geq 0$} \\
\hline
 & $Y_\mathrm{p}+Y_\mathrm{D}$ & $Y_\mathrm{p}$+CMB & $Y_\mathrm{p}+Y_\mathrm{D}$+CMB  & $Y_\mathrm{p}+Y_\mathrm{D}$ & $Y_\mathrm{p}$+CMB & $Y_\mathrm{p}+Y_\mathrm{D}$+CMB  \\
\hline
\multicolumn{7}{|c|}{Electromagnetically coupled DM} \\
\hline
Real scalar &  0.34 & 0.46 & 4.16 &  0.34 & 0.43 & 0.62 \\
Complex scalar & 0.46 & 3.51 & 7.94  &  0.48 & 1.78 & 4.25  \\
Majorana Fermion &  0.46 & 3.19 & 7.76  &  0.47 & 1.52 & 4.18  \\
Dirac Fermion &  0.72  & 6.42 & 10.99  & 0.77 & 4.17 & 8.96 \\
\hline
\multicolumn{7}{|c|}{Neutrino coupled DM} \\
\hline
Real scalar & 1.12 & 11.76 & 8.06 & 1.25 & 13.91 & 9.24\\
Complex scalar & 2.85 & 13.64 & 11.03 & 4.56 & 15.81 & 12.92 \\
Majorana Fermion & 2.53 & 13.50 & 10.79 &  3.92 & 15.31 & 12.34  \\
Dirac Fermion & 4.81  &16.14  & 13.63  &  5.29 &18.34 & 15.09 \\
\hline
\end{tabular}}
\end{table*}
%%%%%%%%%%%%%%

\begin{table*}
\caption{\textbf{The projected lower bounds on DM mass} $m_{\chi}$ at 95\% CL, in units of MeV, for Simons Observatory (SO) and CMB-S4 (S4) (both including data from \textit{Planck}), and their combination with the current primordial abundance measurements. Results are shown for the absence of other relativistic species ($\Delta N_{\nu}=0$) and allowing for their presence ($\Delta N_{\nu}\geq 0$, where $\Delta N_{\nu}$ is marginalized over). We set SBBN values of $N_\mathrm{eff}$ and $Y_{p}$ as the fiducial models in this analysis. \label{tab:forecast}}
\setlength{\tabcolsep}{2.5mm}{
\begin{tabular}{|l|c|c|c|c|c|c|c|c|}
\hline
\multicolumn{1}{|c|}{} & \multicolumn{4}{c|}{$\Delta N_{\nu}=0$} & \multicolumn{4}{c|}{$\Delta N_{\nu}\geq 0$} \\
\hline
 & SO & S4 & $Y_\mathrm{p}+Y_\mathrm{D}$+SO & $Y_\mathrm{p}+Y_\mathrm{D}$+S4 & SO & S4 & $Y_\mathrm{p}+Y_\mathrm{D}$+SO & $Y_\mathrm{p}+Y_\mathrm{D}$+S4 \\
\hline
\multicolumn{9}{|c|}{Electromagnetically coupled DM} \\
\hline
Real scalar & 9.28 & 10.02 & 9.75 & 10.32 & 0.76 & 0.97 & 2.89 & 3.79 \\
Complex scalar & 11.72 & 12.33 & 12.40 & 12.94 & 1.13 & 2.10 & 7.61 & 7.94 \\
Majorana Fermion & 11.63 & 12.29 & 12.34 & 12.84 & 1.07 & 1.97 & 7.44 & 7.69 \\
Dirac Fermion & 13.70 & 14.42 & 14.59 & 15.01 & 2.39 & 5.29 & 10.73 & 10.98 \\
\hline
\multicolumn{9}{|c|}{Neutrino coupled DM} \\
\hline
Real scalar & 9.74 & 10.55 & 8.05 & 9.86 & 10.39 & 11.23 & 10.62 & 11.81 \\
Complex scalar & 12.06 & 12.83 & 11.01 & 12.14 & 12.71 & 13.44 & 13.12 & 13.95 \\
Majorana Fermion & 12.02 & 12.74 & 10.76 & 12.08 & 12.65 & 13.34 & 12.95 & 13.81 \\
Dirac Fermion & 14.27 & 15.03 & 13.06 & 14.01 & 14.91 & 15.80 & 15.77 & 16.62 \\
\hline
\end{tabular}}
\end{table*}
%%%%%%%%%%%%%%

We note that the CMB bound on $m_\chi$ is primarily driven by the measurement of $N_\mathrm{eff}$. In turn, the $N_\mathrm{eff}$ is primarily constrained by \textit{Planck}, with a small contribution from the current ACT and SPT data.
However, the small shifts in the values of $N_\mathrm{eff}$ and $Y_\mathrm{p}$, preferred by different CMB experiments, can notably affect the resulting CMB bounds on $m_\chi$.

For example, for neutrino coupled DM, ACT has a similar constraining power as \textit{Planck}, and the combination of ACT, SPT, and \textit{Planck} strengthens the \textit{Planck} bounds by a factor of $1.4$--$1.8$, depending on the DM spin statistic. This leads to the lower limits between 11 MeV and 16 MeV, at 95\% CL, reported in Table \ref{tab:CMB}. Conversely, SPT has a similar constraining power as \textit{Planck} for electromagnetically coupled DM, but the addition of ACT and SPT \textit{weakens} the bound by up to 30\% for complex scalars and Fermions. For the case of a real scalar, neither \textit{Planck} nor the combination of all CMB data can rule out any DM masses, and SPT only rules out masses below $\sim$2
MeV.

Assuming the presence of additional relativistic degrees of freedom allows for $\Delta N_\nu>0$; this additional freedom can offset the reduction in the value of $N_\mathrm{eff}$ caused by electromagnetically coupled DM, putting small $m_\chi$ in agreement with CMB measurements (as discussed in Sec.~\ref{sec:BBN}). When we fit for $\Delta N_\nu>0$ as a free parameter in our CMB likelihood analysis, \textit{Planck}, ACT, SPT and their combination are consistent with a far broader range of DM masses, down to hundreds of keV.
Finally, letting $\Delta N_\nu>0$ vary as a free parameter generally has comparably smaller effect on the inference of neutrino--coupled--DM mass. Details are listed in Table \ref{tab:CMB}.

%%%%%%%%%%%%%%%%%%%%%%%%%%%%%%%%%%%%%%%
\subsection{Bounds from primordial abundances only}\label{y_bounds}

We next consider only the measurements of primordial helium--4 and deuterium abundances, inferred from the spectra of absorption systems from Ref.~\cite{2020PTEP.2020h3C01P}, and shown as grey bands in Figs.~\ref{fig:Yp} and \ref{fig:D}. Employing the chi--squared method described earlier in this Section, we obtain results presented in Table \ref{tab:CMB_Yp_YD} (columns denoted as ``$Y_\mathrm{p}+Y_\mathrm{D}$''). We find that the primordial abundance measurements alone typically place a significantly weaker lower bound on $m_{\chi}$ than the CMB data, giving $m_\chi\gtrsim 0.3$ MeV, for electromagnetically coupled DM, and $m_{\chi}\gtrsim 1.1$ MeV, for neutrino coupled DM.
Letting $\Delta N_\nu>0$ vary as a free parameter generally has a negligible effect on the primordial--abundance bounds on electromagnetically coupled DM mass, while the bounds on neutrino coupled DM mass are improved.

\begin{figure}[h]
\includegraphics[width=0.4\textwidth]{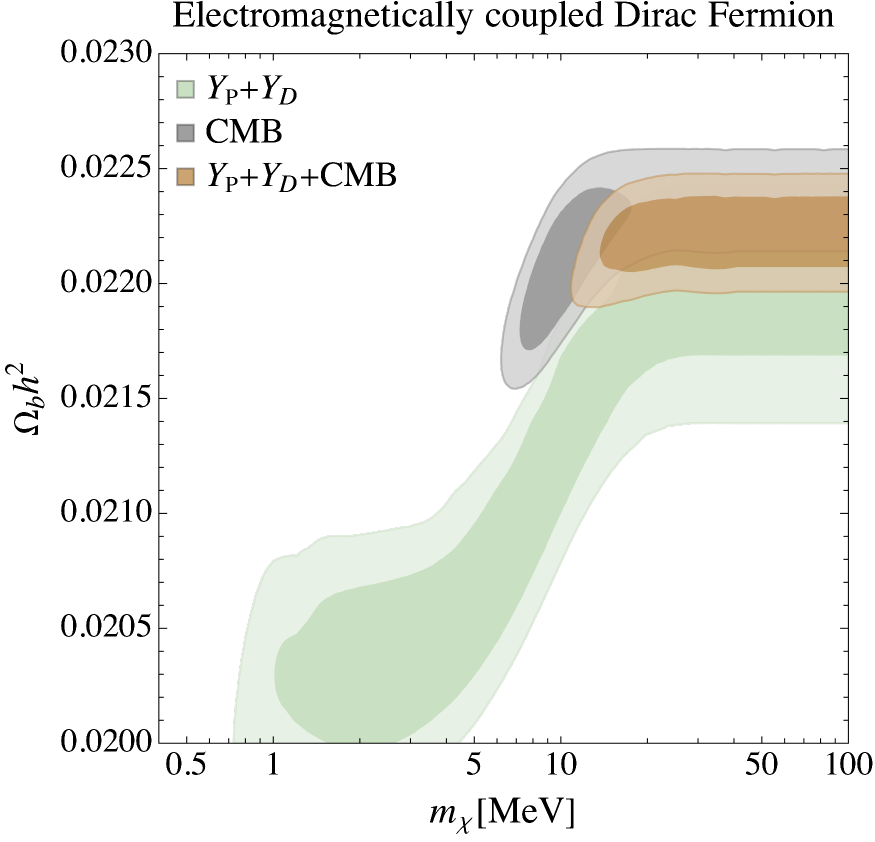}\\
\includegraphics[width=0.4\textwidth]{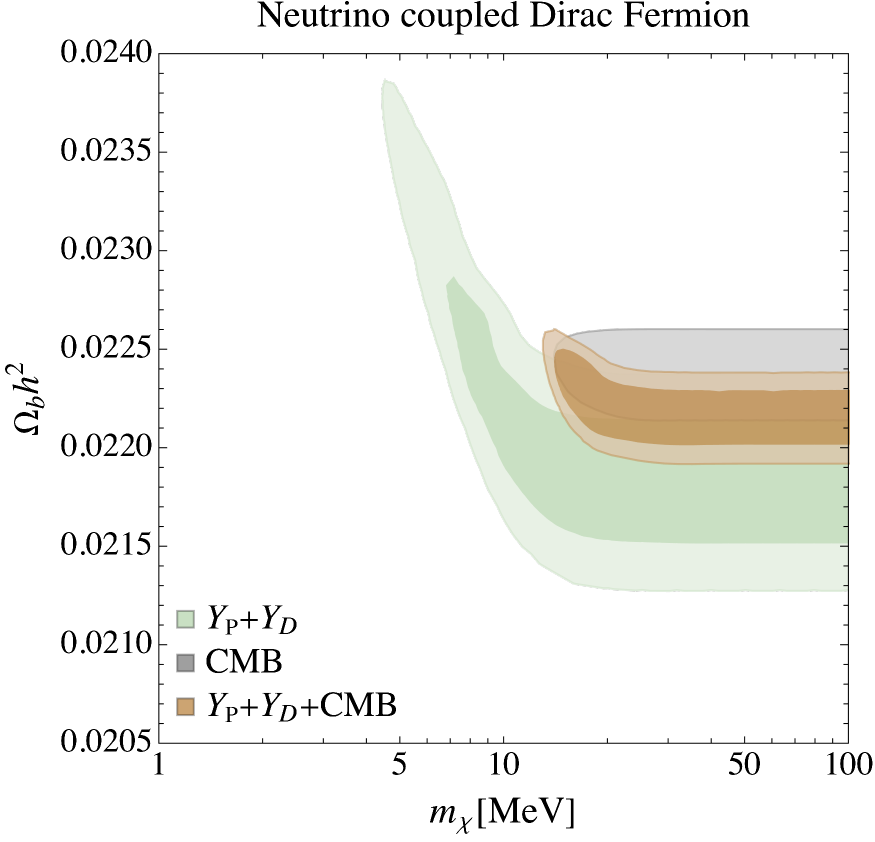}
\caption{The $68\%$ and $95\%$ CL contours on the DM mass and baryon density, for electromagnetically coupled (upper panel) and neutrino coupled (lower panel) Dirac Fermion. The contours are derived using primordial abundance measurements of helium--4 and deuterium only (green), CMB data only (gray; including the measurements from \textit{Planck}, ACT and SPT), and from a joint analysis of the CMB and primordial abundance data (brown). }\label{fig:Yp_Planck}
\end{figure}

We further find a small difference in the preferred value of the baryon density $\Omega_bh^2$ between the primordial abundance measurements and CMB results, shown in Fig.~\ref{fig:Yp_Planck}, consistent with the results of Ref.~\cite{2020JCAP...01..004S, 2021MNRAS.502.2474P}. Specifically, measurements of primordial element abundances prefer a lower value of $\Omega_bh^2$, driven by the measurement of $Y_\mathrm{D}$ (since $Y_\mathrm{p}$ has very little sensitivity to $\Omega_bh^2$). $\Omega_bh^2$ is degenerate with low values of $m_\chi$, which will affect the mass bounds reported here. However, since the values of $\Omega_bh^2$ measured from the CMB and the primordial abundances are consistent at the $3\sigma$ level, we proceed to combine all the available data in the following. 
%%%%%%%%%%%%%%

%%%%%%%%%%%%%%%%%%%%%%%%%%%%%%%%%%%%%%%
\subsection{Joint bounds}\label{sec:joint_bounds}

%%%%%%%%%
\begin{figure}
\includegraphics[width=0.37\textwidth]{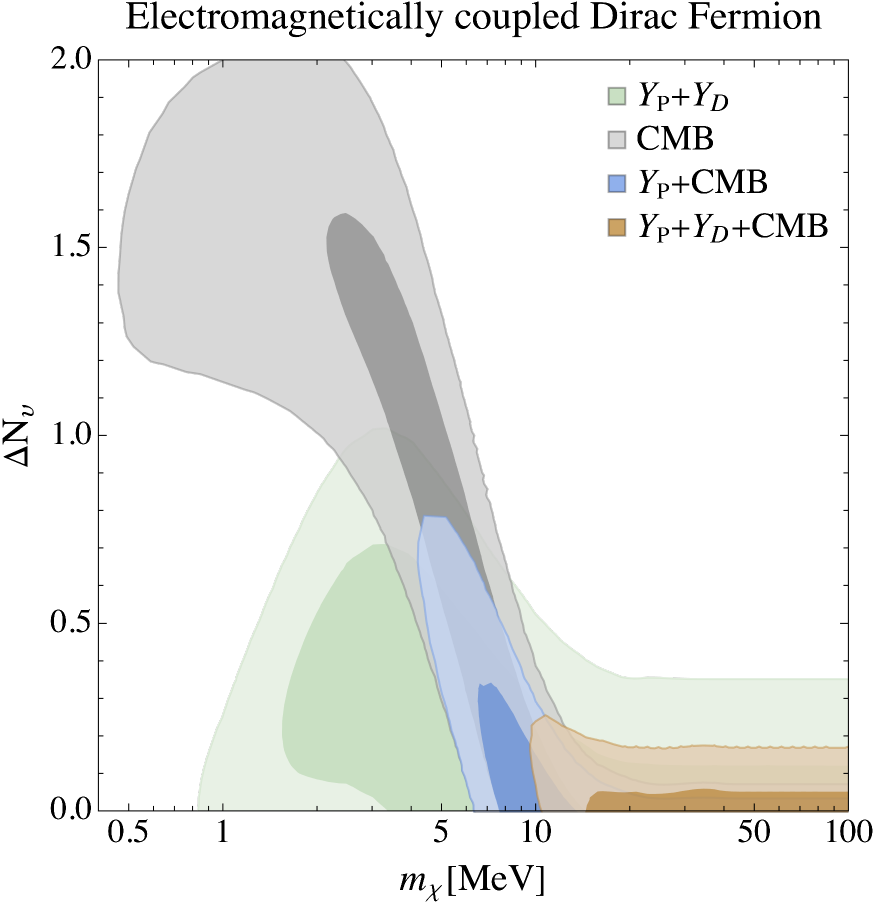}\\
\includegraphics[width=0.37\textwidth]{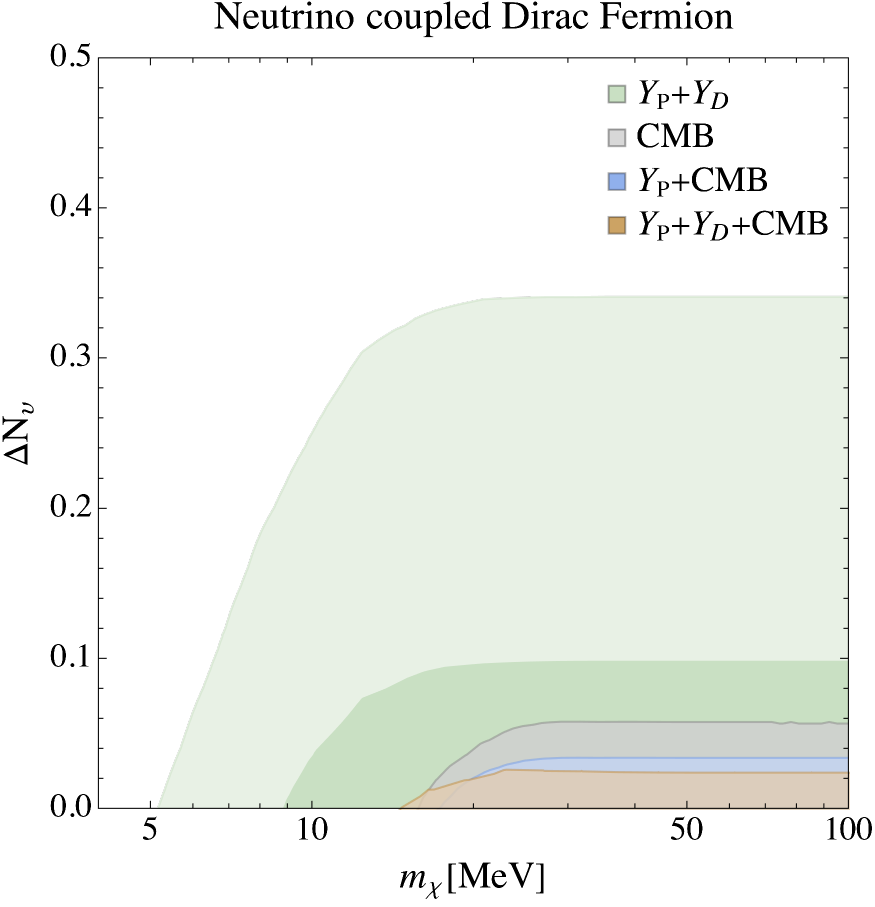}
\caption{The $68\%$ and $95\%$ CL contours on the DM mass and the effective number of additional relativistic species, for electromagnetically coupled (upper panel) and neutrino coupled (lower panel) Dirac Fermion DM. These results are derived using primordial abundance measurements of helium--4 and deuterium only (green), CMB data only (grey; including  the measurements from \textit{Planck}, ACT, and SPT), and using joint analyses of CMB data and the primordial abundance measurements (with helium--4 only: blue, with both helium--4 and deuterium: brown). Note that $\Omega_b h^2$ is marginalized over.}\label{fig:CMB_Yp_YD}
\end{figure}
%%%%%%%%
%%%%%%%%%%%%%%
\begin{figure}[h]
\includegraphics[width=0.38\textwidth]{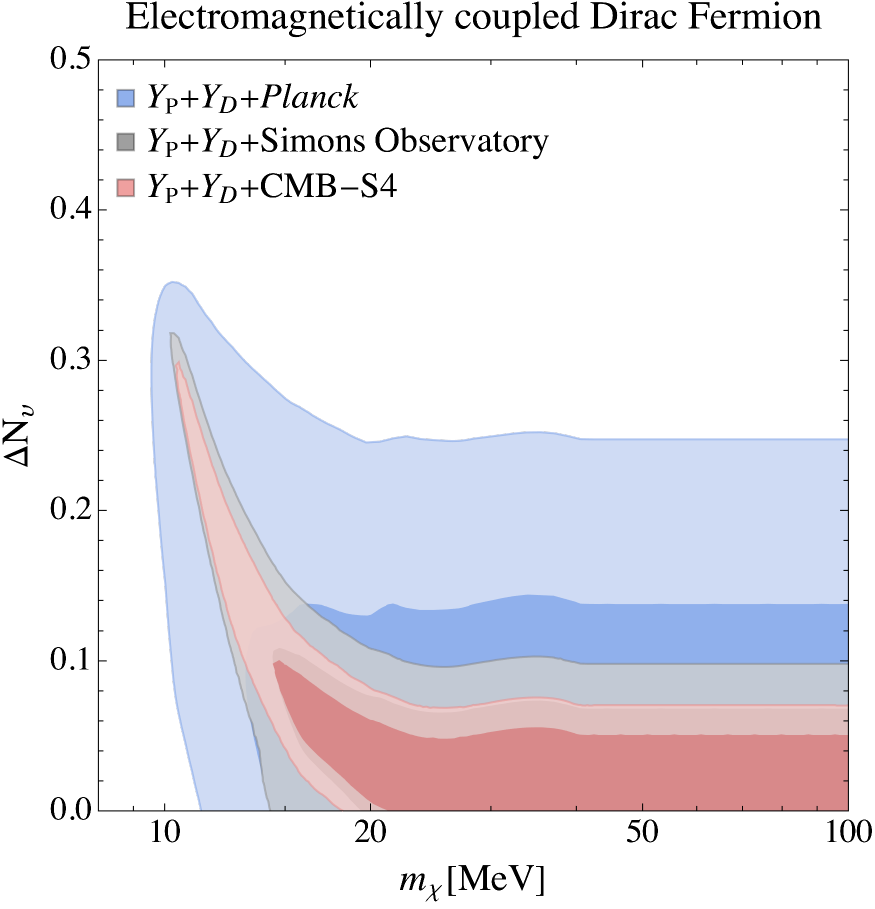}\\
\includegraphics[width=0.38\textwidth]{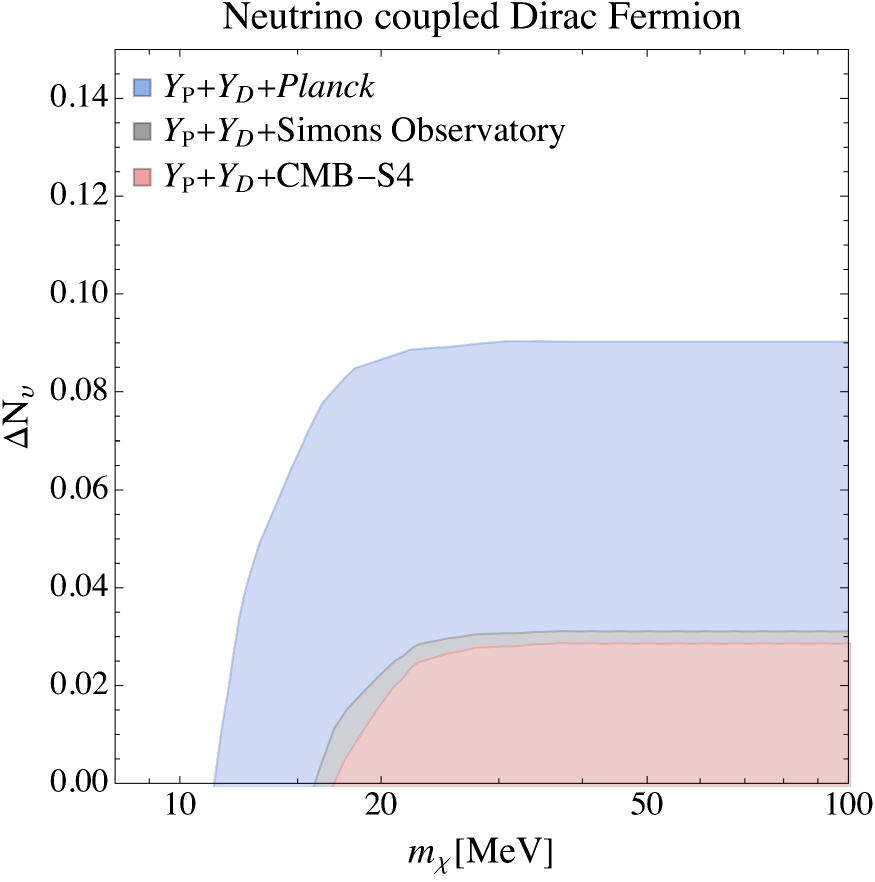}
\caption{The current and future $68\%$ and $95\%$ CL contours on the DM mass and the effective number of additional relativistic species, for electromagnetically coupled (upper panel) and neutrino coupled (lower panel) Dirac Fermion DM. The contours are derived assuming joint analyses of current primordial abundance measurements with CMB data from \textit{Planck} (blue), Simons Observatory (grey), and CMB-S4 (red), allowing for the presence of equivalent neutrinos $\Delta N_{\nu}\geq 0$. Note that $\Omega_b h^2$ is marginalized over.}\label{fig:forecast}
\end{figure}
%%%%%%%%%%%%%%

When no additional relativistic species are allowed, the addition of $Y_\mathrm{p}$ measurement to the combination of the CMB data excludes masses below $\sim$0.5 MeV for electromagnetically coupled real scalar, which is unconstrained by CMB alone. For all other models where the mass bounds from CMB are present, the addition of $Y_\mathrm{p}$ leads to only a small improvement. 
The joint CMB + $Y_\mathrm{p}$ limits on $m_{\chi}$ for all DM models are presented in Table \ref{tab:CMB_Yp_YD}.

We next combine the measurement of  $Y_\mathrm{D}$ with $Y_\mathrm{p}$ + CMB data and show the resulting 95\% lower limits on $m_{\chi}$ in Table \ref{tab:CMB_Yp_YD}. If $\Delta N_\nu=0$ is kept fixed, we see that the mass bound improves for electromagnetically coupled DM, excluding masses below $\sim$4 MeV at 95\% CL, for all DM spin statistics; conversely, the bound relaxes for neutrino coupled DM, excluding masses below $\sim$8 MeV at 95\% CL, for all DM spin statistics.

Allowing $\Delta N_\nu$ to vary as a free parameter leads to the bounds illustrated in Fig.~\ref{fig:CMB_Yp_YD} and in the rest of Table \ref{tab:CMB_Yp_YD}. From the Figure, we see that there is a degeneracy between $m_{\chi}$ and $\Delta N_{\nu}$ at $m_{\chi}\lesssim 20$ MeV, for electromagnetically coupled particles, regardless of the spin statistic. This means that smaller masses are consistent with the data, if new relativistic species are allowed in the cosmological model. In contrast, for neutrino coupled DM, the mass bounds are not very sensitive to the choice of $\Delta N_{\nu}$. 

%%%%%%%%%%%%%%%%%%%%%%%%%%%%%%%%%%%%%%%%%%%
\section{Future constraints}\label{sec:forecasts}

We now quantify the degree to which measurements of small--scale CMB polarization and temperature anisotropy can improve current bounds on $m_\chi$. For this purpose, we consider two future ground--based CMB experiments: Simons Observatory (SO) \cite{2019JCAP...02..056A} and CMB-S4 \cite{2016arXiv161002743A, 2019arXiv190704473A}.  
We next use the forecasted covariance matrices for the three key parameters $\{\Omega_b h^2, N_\mathrm{eff}, Y_\mathrm{p}\}$ reported in Ref.~\cite{2020JCAP...01..004S} for the two CMB experiments\footnote{This reference has only focused on a subset of models discussed in this work.}, computed assuming that both SO and CMB-S4 are combined with the current \textit{Planck} data. The forecasts assume a Gaussian likelihood, such that
\begin{equation}
\chi_{\rm{CMB}}^2  = \left(\textbf{X} - \textbf{X}_{\mathrm{Fiducial}}\right) \textbf{Cov}^{-1} \left(\textbf{X} - \textbf{X}_{\mathrm{Fiducial}}\right)^T, 
\end{equation}
where we assume that the future best-fit cosmology corresponds to standard BBN,
\begin{equation}
    \textbf{X}_{\mathrm{Fiducial}}=\{\Omega_b h^2, N_\mathrm{eff}, Y_\mathrm{p}\} = \{0.02236,\, 3.046,\, 0.2472\}
\end{equation}
and
\begin{align}
\textbf{Cov} = \,
 \left(
\begin{array}{ccc}
\sigma_1^2 & \sigma_1 \sigma_2 \rho_{12} & \sigma_1 \sigma_3 \rho_{13}  \\
\sigma_1 \sigma_2 \rho_{12}     & \sigma_2^2 & \sigma_2 \sigma_3 \rho_{23} \\
\sigma_1 \sigma_3 \rho_{13}     & \sigma_2 \sigma_3 \rho_{23}        & \sigma_3^2 \\
\end{array}
\right) \,, 
\end{align}
where $(\sigma_{1},\sigma_{2},\sigma_{3}) = (0.000073,0.11,0.0066)$ and
$(\rho_{12},\rho_{13},\rho_{23}) = (0.072, 0.33, -0.86)$ for SO, $(\sigma_{1},\sigma_{2},\sigma_{3}) = (0.000047, 0.081, 0.0043)$ and $(\rho_{12}, \rho_{13}, \rho_{23}) = (0.25, 0.22, -0.84)$ for CMB-S4.
We then convert these covariances into 95\% CL intervals for DM mass $m_\chi$, and list the results in Table \ref{tab:forecast}. Fig.~\ref{fig:forecast} shows the projected constraints on $m_\chi$ and $N_\mathrm{eff}$ for Dirac Fermion, when future CMB measurements are combined with the current measurements of primordial element abundances. We find that the addition of primordial element abundance measurements has minimal effect on the projected bounds from future CMB data, even in the case where $\Delta N_\nu$ is a free parameter.

However, if future CMB experiments return the best--fit values of  $\{\Omega_b h^2, N_\mathrm{eff}, Y_\mathrm{p}\}$ that are \textit{inconsistent} with the standard BBN, they might have sufficient accuracy to detect thermal--relic DM mass. For example, if the best--fit parameters for SO are the current best--fit values obtained from a combined analysis of all CMB experiments (\textit{Planck}+ACT+SPT), such measurements will have constraining power to exclude neutrino coupled DM at $>2\sigma$ significance. Equivalently, if thermal--relic DM has mass just below the detection threshold from the current CMB$+Y_\mathrm{p}+Y_\mathrm{D}$ analyses (around 10 MeV for a Dirac Fermion), future CMB experiments will be able to favor light thermal--relic DM at 2--3$\sigma$ confidence and return measurement of its mass with a few MeV accuracy.
%%%%%%%%%%%%%%%%%%%%%%%%%%%%%%%%%%%%%%%%%%%
\section{Summary and discussion}\label{sec:summary}

%%%%%%%%%%%%
\begin{figure}
\includegraphics[width=0.45\textwidth]{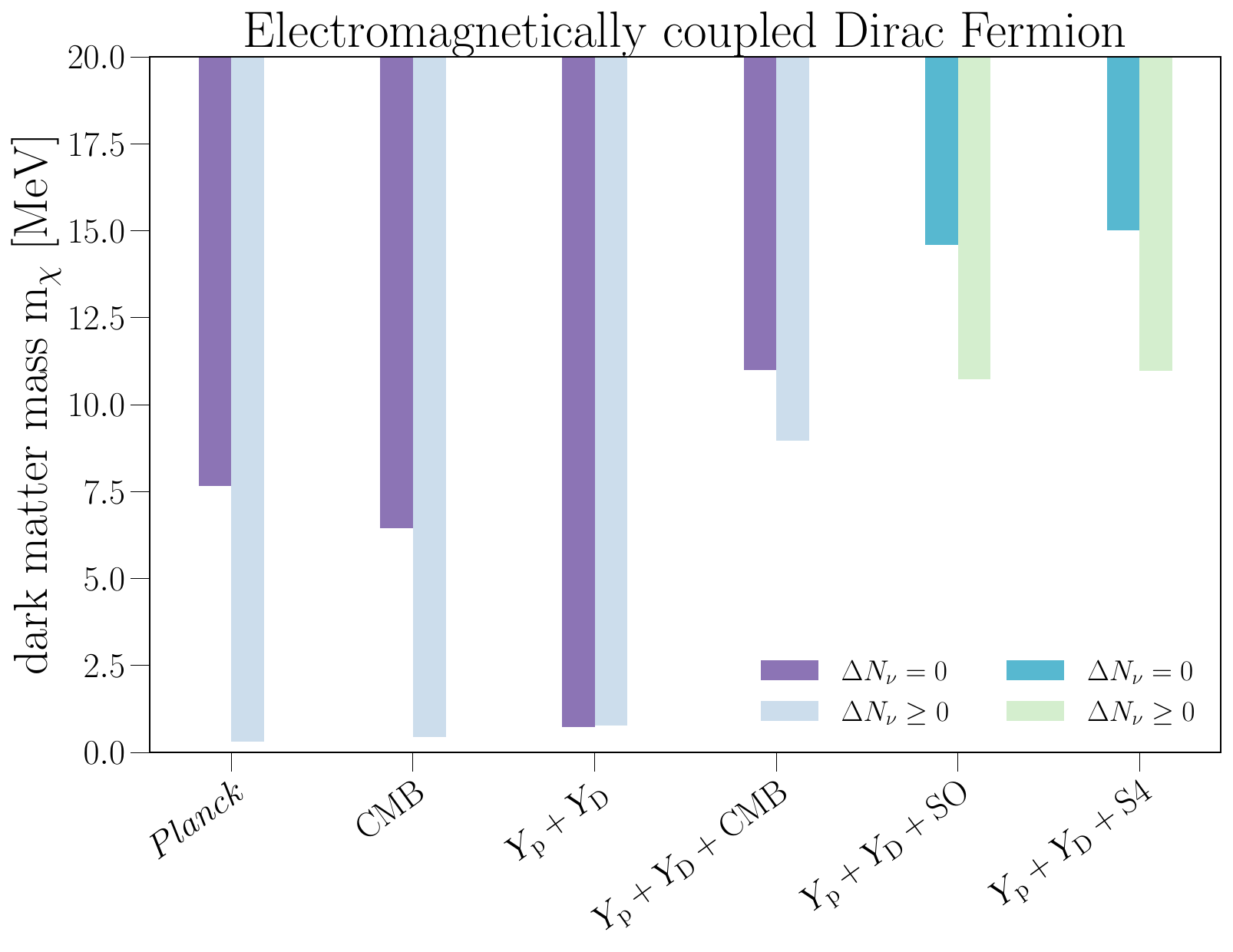}\\ 
\includegraphics[width=0.45\textwidth]{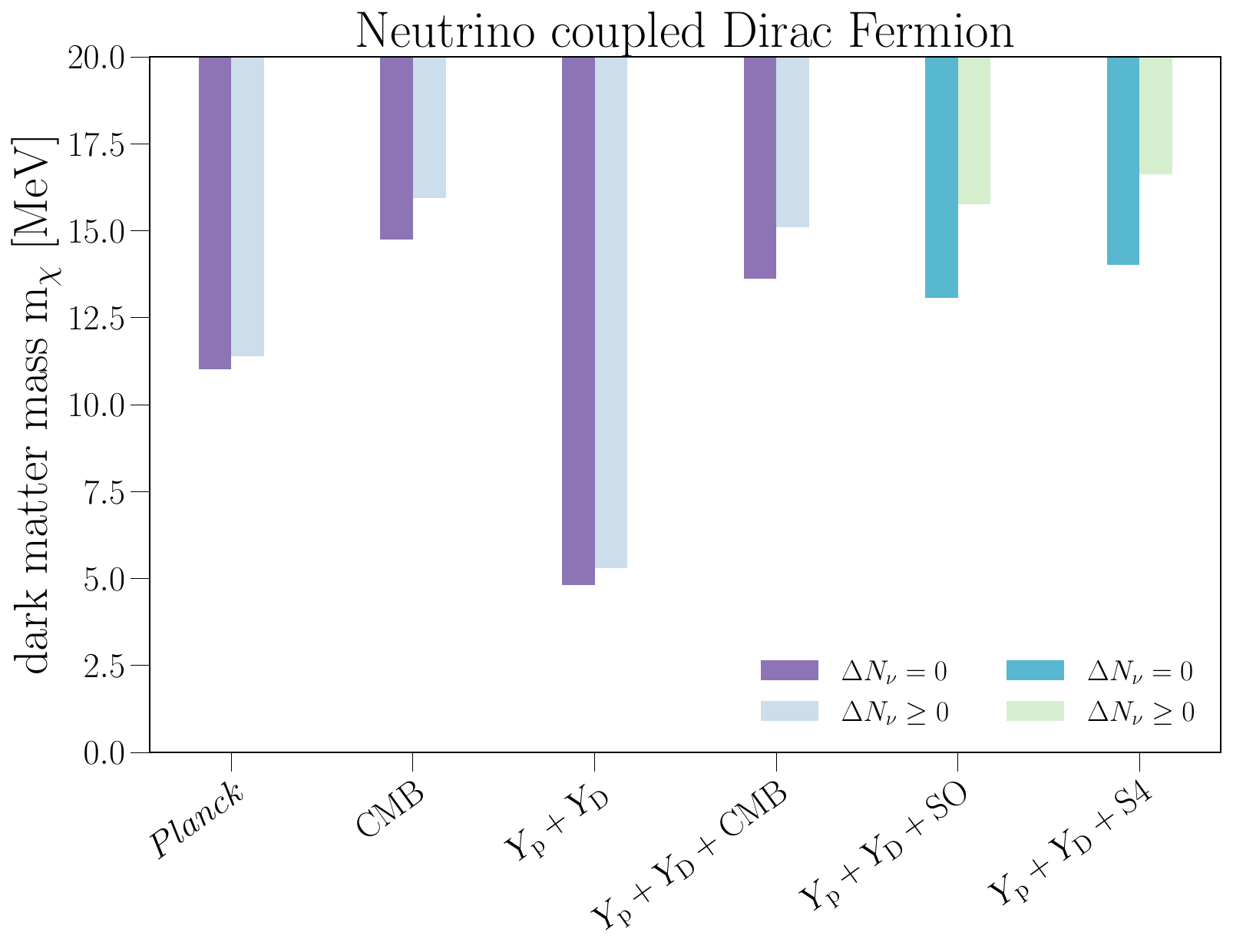}
\caption{Summary of current cosmological bounds on thermal--relic DM mass, for the example of a Dirac Fermion, where the values for the mass consistent with a given data set (at 95\% confidence) correspond to the colored regions. Note that  CMB indicates a joint analysis of all CMB data from \textit{Planck}, ACT and SPT. The purple  shaded regions are constraints derived in Sec.~\ref{sec:constraints} and the forecasts for joint analyses of $Y_\mathrm{p}+Y_\mathrm{D}$+\textit{Planck} with Simons Observatory (SO) and CMB-S4 (S4) are shown as shades of green. The light shaded regions result from analyses that allow additional relativistic species in the universe, and dark shaded regions are from analyses that assume no additional light species. The top and bottom panels represent the case of electromagnetically coupled and neutrino coupled Dirac Fermion DM, respectively. }\label{fig:summary}
\end{figure}
%%%%%%%%%%%%
We have combined \textit{Planck} 2018 data with the most recent public releases of ground--based measurements of the small--scale CMB temperature and polarization anisotropy from ACT DR4 and SPT--3G, and with the measurement of the primordial abundances of helium--4, $Y_\mathrm{p}$, and deuterium, $Y_\mathrm{D}$. We derived the most general bounds on the mass of a light thermally--coupled DM particle that couples to either SM neutrinos or photons, affecting the expansion history and the process of BBN. To quantify the effects of mass $m_\chi$ on CMB anisotropy, our likelihood analyses have included $N_\mathrm{eff}$ and $Y_\mathrm{p}$ as free parameters, in addition to the standard cosmological parameters. We converted the resulting posterior probability distribution on the relevant cosmological parameters to the posterior probability distribution on $m_\chi$, inferring the lower mass limits reported in Tables \ref{tab:CMB} and \ref{tab:CMB_Yp_YD}. We also explored DM mass measurements that might be possible to achieve in the future, with SO and CMB-S4, with key results reported in Table \ref{tab:forecast}. Importantly, we quantified how the presence of additional relativistic species may affect the inference of DM mass, and found a moderate impact on electromagnetically coupled DM models, and a negligible impact on neutrino coupled models.  
A summary of the bounds obtained from various combinations of data is illustrated in Fig.~\ref{fig:summary}, for the representative case of a Dirac Fermion. 

%%%%%%%%%%%%%%%%%%%%%%%%%%
\subsection{Key results}

For DM coupled to neutrinos, combining ACT and SPT with \textit{Planck} improves the lower limit on $m_\chi$ by 40\%--80\%, depending on DM spin statistic. The strongest lower limit on the mass is around 15 MeV, at 95\%CL, regardless of whether $\Delta N_{\nu}$ is fixed to 0 or allowed to vary. The CMB is significantly more constraining than the $Y_\mathrm{p}$ and $Y_\mathrm{D}$ measurements alone. 

For electromagnetically coupled DM, the addition of ACT and SPT to \textit{Planck} weakens previous bounds by $\lesssim 30$\% for $\Delta N_{\nu}=0$. This occurs mainly because the joint CMB data prefer a lower helium abundance $Y_\mathrm{p}$ and a slightly higher  value  of $N_\mathrm{eff}$ than \textit{Planck} alone. We emphasize that these shifts have small statistical significance, and the standard BBN is still within 2$\sigma$ of the best--fit value for the combined CMB measurements.
Allowing additional light species also significantly weakens CMB bounds on electromagnetically coupled models, resulting in the change of the lower limit on the mass from a few MeV to hundreds of keV. In this case, the mass of a Majorana Fermion is only constrained by the CMB when all three CMB data sets are combined. 
Finally, when $\Delta N_{\nu}>0$, the primordial abundance measurements and the CMB are highly complementary in their constraining power, and the mass bounds are improved by a factor of a few when the two sets of data are combined; the lower limit on the mass is between $\sim600$ keV and $\sim9$ MeV, depending on DM spin statistic.

%%%%%%%%%%%%%%%%%%%%%%%%
\subsection{Robustness of the current measurements}

We highlight the small differences in the preferred values of cosmological parameters found between different data sets considered in this study, and discuss their effects on the current bounds on thermal--relic DM. First, as discussed in Sec.~\ref{sec:constraints}, the addition of $Y_\mathrm{D}$ measurement to the CMB \textit{worsens} most of the CMB--only bounds on neutrino coupled DM because of the mild tension between the $\Omega_bh^2$ values inferred from the two data sets \cite{2020JCAP...01..004S,2021MNRAS.502.2474P}. This tension is currently at $\sim1.8\sigma$ level, but future high--precision spectral characterization of damped Lyman--$\alpha$ systems at high redshift can lead to further improvement in the measurement of $Y_\mathrm{D}$ \cite{2013arXiv1310.3163M}, while the next-generation CMB measurements from SO and CMB--S4 will significantly decrease the uncertainty on the measurement of the baryon fraction as well. If the tension persists and increases in significance with future data, neutrino coupled DM model will be insufficient to describe all data, and new physics may be required to explain this discrepancy and infer bounds on DM mass accurately.

Secondly, as discussed in Sec.~\ref{sec:constraints}, the mass bounds for all DM models shift when ground--based CMB measurements are combined with \textit{Planck}, because different CMB experiments prefer slightly different values of $N_\mathrm{eff}$. Notably, ACT is only consistent with the standard BBN within 2$\sigma$, preferring a slightly lower $N_\mathrm{eff}$, as discussed in Ref.~\cite{2020JCAP...12..047A}; at the same time, SPT prefers a larger value than the \textit{Planck} best--fit. This mild discrepancy can be a statistical fluctuation, or a consequence of a systematic effect. Preliminary checks we performed in our analysis do not point directly to either, and a more detailed investigation is needed to understand this difference. Notably, the preference towards lower value seen in ACT is \textit{not} obviously driven by low multipoles in EE and TE power spectra that drive preference for other beyond--$\Lambda$CDM models \cite{2021arXiv210904451H}. Future data sets from ACT and those from SO and CMB--S4 will be able shed more light on this issue and recover a more robust bound on DM mass. 

%%%%%%%%%%%%%%%%%%%%%%%%%%%%%%%%%%%%%%%%%%%%%%%%%
\subsection{The potential of future CMB measurements}

Key science goals of Simons Observatory and CMB-S4 include high--precision measurements of $N_\mathrm{eff}$ and $Y_\mathrm{p}$. As discussed in Sec.~\ref{sec:forecasts}, if these results confirm the standard BBN scenario, the bounds on thermal--relic DM will improve by several MeV,  as compared to current analyses for most models\footnote{Due to the shift in the preferred values of $N_\mathrm{eff}$ and $Y_{p}$, the combination of all CMB data can provide tighter constraints than the forecast bounds for neutrino coupled DM in some scenarios.}, excluding most of the mass range that could affect the process of BBN.

Alternatively, in the scenario where future CMB measurements converge around the current best--fit values of $N_\mathrm{eff}$ and $Y_\mathrm{p}$ from \textit{Planck}+ACT+SPT, both SO and CMB--S4 will have sufficient sensitivity to exclude standard BBN at several $\sigma$; if the DM mass is just above the current upper limit, these experiments will be able to \textit{measure} it with a few--MeV accuracy.

In the end, it is important to note that the effects of neutrinophilic and electrophilic DM on cosmological observables can mutually cancel, and may be obscured by the presence of other new physics. If reality is not well described by standard cosmology and standard BBN, many of the mass bounds derived in this and similar studies will require revision.

\section*{Acknowledgements}
The authors acknowledge helpful conversations with Jo Dunkley, related to testing the results of ACT analysis. RA and VG acknowledge the support from NASA through the Astrophysics Theory Program, Award Number 21-ATP21-0135. EC acknowledges support from the STFC Ernest Rutherford Fellowship ST/M004856/2, STFC Consolidated Grant ST/S00033X/1and from the European Research Council (ERC) under the European Union’s Horizon 2020 research and innovation programme (Grant agreement No. 849169). JCH acknowledges support from NSF grant AST-2108536. The Flatiron Institute is supported by the Simons Foundation.

\bibliographystyle{IEEEtran}
\bibliography{wimp}

\end{document}